\documentclass[preprint]{aastex63}

\received{\today}
\revised{\today}
\accepted{\today}
\submitjournal{ApJS}

\shorttitle{Rotation periods of \textit{K2}C9 asteroids}
\shortauthors{Podlewska-Gaca et al.}

\begin{document}

\title{Determination of rotation periods for a large sample of asteroids from \textit{K2} Campaign 9}

\correspondingauthor{Edyta Podlewska-Gaca}
\email{edypod@amu.edu.pl}

\author[0000-0003-3790-3847]{Edyta Podlewska-Gaca}
\affiliation{Astronomical Observatory Institute, Faculty of Physics, Adam Mickiewicz University,\\ ul. Słoneczna 36, 60-286 Poznań, Poland}

\author[0000-0002-9245-6368]{Rados\l{}aw Poleski}
\affiliation{Astronomical Observatory, University of Warsaw, Al. Ujazdowskie 4, 00-478 Warszawa, Poland}

\author[0000-0002-3466-3190]{Przemys\l{}aw Bartczak}
\affiliation{Astronomical Observatory Institute, Faculty of Physics, Adam Mickiewicz University,\\ ul. Słoneczna 36, 60-286 Poznań, Poland}

\author[0000-0003-0356-0655]{Iain McDonald}
\affiliation{Jodrell Bank Centre for Astrophysics, Alan Turing Building, University of Manchester, Manchester, M13 9PL, UK}
\affiliation{Open University, Walton Hall, Kents Hill, Milton Keynes, MK7 6AA, UK}

\author[0000-0001-5449-2467]{Andr\'as P\'al}
\affiliation{Konkoly Observatory, Research Centre for Astronomy and Earth Sciences, H-1121 Budapest, Konkoly Thege Mikl\'os \'ut 15-17, Hungary}
\affiliation{E\"otv\"os Lor\'and University, H-1117 P\'azm\'any P\'eter s\'et\'any 1/A, Budapest, Hungary}
\affiliation{MIT Kavli Institute for Astrophysics and Space Research, 70 Vassar Street, Cambridge, MA 02109, USA}

\begin{abstract}
Kepler mission is a powerful tool in the study the different types of astrophysical objects or events in the distant Universe. However, the spacecraft gives also the opportunity to study Solar System objects passing in the telescope field of view. 
The aim of this paper is to determine for the first time the rotation periods of a number of asteroids observed by the \textit{Kepler} satellite during the \textit{K2} Campaign 9.
 Using publicly available data from Kepler mission we have used the Modified Causal Pixel Model (MCPM) code to perform the aperture-like and PRF-like photometry of 1026 asteroids. 
 The results allowed us to determine the rotation periods for 188 asteroids. For asteroids with rotation periods previously measured, we compared the results and found very good agreement. 
There are additional 20 asteroids for which we obtained lower limits on rotation periods and in all cases these limits are at least $100~\mathrm{h}$.

\end{abstract}


\section{Introduction}

Asteroids are thought to be a planetesimal leftover from the formation stage of
our Solar System and therefore carry information on its early state and
evolution. Since 1801 and the discovery of the first known asteroid -- (1)
Ceres -- our understanding of the processes underlying the Solar System's
creation and evolution has increased tremendously. So far, we have detected more
than $8\times 10^5$ asteroids, but for most of them the visual brightness and
orbit are the only known parameters (For details see Minor Planet Center\footnote{\url{https://minorplanetcenter.net/}}).

Rotation period ($P$) is one of the most important physical parameters of an asteroid.
For example, it allows the 
computation of the rotation period phase for an asteroid from sparse in-time absolute magnitude measurements and the creation of phase curves, that are corrected for lightcurve amplitude.
From those, geometric albedo and taxonomic classes can be deduced giving information on an asteroids' chemical composition \citep{Shevchenko}. The spin and shape modelling of asteroids also requires precise knowledge of the rotational period \citep{Kas}. The volume and the density determination becomes possible when photometrically derived rotation period is combined with other techniques like adaptive optics imaging \citep[e.g.,][]{Vernazza, Carry}, occultation chord timings \citep[e.g.,][]{Durech2011}, or thermal infrared observations \citep[e.g.,][]{Hanus,paperI}. Only a handful of Solar System objects have been visited by spacecrafts giving us a more or less complete picture of their structure:  e.g. (21) Lutetia pictured during the Rosetta mission flyby, or (101955) Bennu studies by OSIRIS-REx \citep{Rosetta, OSIRIS}. Decades of precise photometric observations and campaigns resulted in the determination of physical parameters for more than 2000 objects \citep[Database of Asteroid Models from Inversion Techniques, DAMIT;][]{Durech2010}\footnote{\url{http://astro.troja.mff.cuni.cz/projects/damit}}. The rotational periods are currently known for about 33,000 asteroids \citep{Warner}. Nonetheless, our understanding of asteroids is far from complete and the rate of discovering new objects is still much faster than the rate of studying them.

Precise determination of $P$ requires a lightcurve with at least 100 data points that cover a few rotation periods. This is very difficult to fulfil, especially for fainter objects, for which large-aperture telescopes are required. Moreover, for objects with long rotation periods, even the whole night is not enough in order to cover the rotation period fully. Rotation periods can be measured best if one uses long and continuous time-series photometry. In recent years, this has been possible by using the \textit{Kepler} or \textit{Transiting Exoplanet Survey Satellite} (TESS) satellites \citep{Borucki, Ricker}.

The original field observed by \textit{Kepler} was far from the ecliptic \citep{Borucki}, and Solar System bodies could not be observed. After a failure of the two of its four reaction wheels, the mission was re-purposed as \textit{K2} \citep{Howell} and used solar radiation pressure to keep the telescope stable. During the \textit{K2} mission, the telescope could only be oriented such that it observed close to the ecliptic plane, which allowed to detect a large number of Solar System objects. In this work, we measured rotation periods of 188 asteroids by analyzing time-series photometry from the \textit{K2} Campaign 9 (\textit{K2}C9).

The \textit{Kepler} camera has more pixels than the number available for downlink. Hence, both in \textit{Kepler} and \textit{K2} missions for most targets only the postage stamps were selected for downlink \citep{Bryson10}. In such cases detection of asteroids is possible \citep{Szabo15}, but does not allow continuous time-series photometry to be obtained. There were a few exceptions where larger continuous superstamps were selected in order to collect photometry of stellar clusters, Uranus, Neptune, and the Galactic bulge. In the previous studies of Main Belt Asteroids (MBA) in the \textit{K2} superstamps, \citet{Szabo16} obtained photometry of 1020 objects and measured 37 rotation periods, while for \citet{Molnar18} study these numbers were 608 and 90, respectively. \citet{Pal18} lists a number of other investigations of Solar System objects which used \textit{K2} data and further analysis were published \citep[e.g.,][]{Marton20}. Next, using the K2 data \cite{Szabo20} investigated the rotational properties of Hilda group and \cite{Kalup} explored the Trojan asteroids.
The \textit{K2} mission has ended and currently \emph{TESS} satellite collects high-cadence photometry, which allows further studies of Solar System objects \citep[e.g.,][]{Pal_tess}.

The goal of the \textit{K2}C9 was to observe a large superstamp covering an almost continuous area of the Galactic bulge, in order to observe microlensing events simultaneously from \textit{Kepler} and Earth \citep{Gould13,Henderson16}. The \textit{K2}C9 superstamp covers $3.7~\mathrm{deg^2}$ in total \citep[see Figure~7 in][]{Henderson16} and the ecliptic latitudes vary between $-5\fdg9$ and $-3\fdg5$. The combination of large sky area monitored, almost un-interupted observations, and location close to Ecliptic allow observations of a large number of asteroids and, hence, determination of their rotation periods. Thanks to the campaign length of 71 days, the rotation periods can be measured even for slowly rotating asteroids. Efficient extraction of the photometry of the \textit{K2}C9 asteroids is complicated by the extremely high stellar density of the Galactic bulge, hence, the \textit{K2}C9 data were deemed unsuable for asteroid research \citep{Molnar18}. We overcome these problems by using Modified Causal Pixel Model \citep[MCPM;][]{Poleski_method} developed for photometry of microlensing events.

As a result of our study, we have determined for the first time rotation periods for more than 150 asteroids, and confirmed periods for 37 objects, which were used as test cases for our method. Moreover, we provide here the uncertainties of the determined periods.

The paper is organised as follows. In the next section we describe the \textit{K2}C9 observations and photometry. In the Section 3 we present the method of period and its uncertainty determination, and in Section 4 we show the results. Conclusions are given in Section 5.

\section{\textit{K2} observations and photometry extraction}

The \textit{Kepler} satellite is equipped with a  $0.95~\mathrm{m}$ diameter mirror and a mosaic CCD camera with a field of view of $115~\mathrm{deg^2}$. The pixel scale of the CCD camera is $3\farcs98$. The integration time of a single exposure was $30~\mathrm{min}$ during \textit{K2}C9. The telescope pointing drifts with an amplitude of around 1 px and thrusters are fired every $\approx6~\mathrm{h}$ in order to compensate for the drift. The exposures taken during thrusters firing are not usable. The camera is equipped with a wide-band optical filter.

The \textit{K2}C9 goal was to measure microlensing parallaxes \citep{Henderson16}, hence, a sky-area with the highest microlensing rate was targeted, i.e., the central part of the Galactic bulge. There are up to three stars brighter than $I=20~\mathrm{mag}$ per \textit{K2} pixel \citep{Henderson16}, which combined with telescope pointing drift results in an extremely challenging dataset. In order to increase the sky-area observed, the campaign was split into two subcampaigns with a data downlink in between. The time ranges of subcampaigns C9a and C9b were JD $2,457,501.1$--$2,457,527.4$ (22 April 2016--18 May 2016), and $2,457,531.1$--$2,457,572.4$ (22 May 2016--2 July 2016), respectively. There are gaps in the time series of asteroids due to break between subcampaigns and in many cases asteroids moving out of and back into the superstamp. 

The list of objects to be studied was compiled from two sources. First, McDonald et al. (submitted) searched for short-lasting microlensing events in the \textit{K2}C9 data and in their verification process, they compiled a list of 1149 asteroids by running the ISPY service by JPL/NASA in steps of $2.5~\mathrm{d}$.
This list is limited to objects brighter than $21~\mathrm{mag}$. Second, we searched for asteroids using the EPHEMD utility which has been designed to handle the MPC object database in an offline manner in order to execute bulk searches for asteroids on large field-of-view optical telescopes \citep[see also][]{Molnar18,Pal18,Pal_tess} as well as to query serendipitous detections on other optical setups \citep[see e.g.,][]{Szakats20}. The \textit{K2}C9 superstamp falls in five channels of the camera and we queried the central channel of the superstamp,
which resulted in 461 asteroids. These two lists partly overlap, hence, we have 1395 unique objects. We obtained ephemerides of these objects using the Horizons service by JPL/NASA. It turned out that some of the selected objects passed close to the field but have not passed through the field of view or passed the field of view during the break between subcampaigns. We are left with 1026 asteroids which were observed by \textit{Kepler} during the \textit{K2}C9 at least once. These objects are
mainly MBA, but there are also examples of Jupiter Trojans (18071, 13229, 141577, 228119, and 2009 WJ58), a Near Earth Object (NEO) from the Apollo group (317643, 485823, and 2012 YY6), and a Mars-crossing asteroid (391595). The \textit{K2} images were downloaded from the Mikulski Archive for Space Telescopes\footnote{\url{https://archive.stsci.edu/k2/}}. We interpolated the ephemerides to BJD epochs for which \textit{K2}C9 images were acquired. Due to the large field of view of the \textit{Kepler} camera, the BJD epochs 
of a single exposure are different for different parts of the camera. We interpolate the ephemerides to the BJD epoch provided for a part of the camera at which given asteroid is observed.

To extract the photometry we use the MCPM \citep{Poleski_method} approach, which is a significantly modified version of the Causal Pixel Model (CPM) by \citet{Wang16}. The CPM uses linear correlations of signals observed in different parts of the camera to remove the trends produced by a combination of blending and telescope pointing jitter. The goal of the CPM was to find planetary transits (which manifest as low-amplitude and short-lasting signals) in sparse fields. The MCPM was designed to extract photometry of microlensing signals (which are long and large amplitude) from the \textit{K2}C9 data. The most important changes relative to CPM are combination of signals from a few pixels, weighting these signals using the \textit{Kepler} Point Response Function \citep[PRF;][]{Bryson10}, and calculating residuals based on a full model of the astrophysical signal. 
In MCPM, the flux of each pixel is represented as a sum of instrumental trends (equal to fluxes in other pixels multiplied by linear coefficients) and astrophysical flux weighted by the PRF. In the final step of MCPM, the astrophysical flux from a few pixels is combined into a final estimate of the astrophysical flux.
This final step can be done in two ways: just summing the fluxes from a few pixels (aperture-like photometry) or weighting the fluxes using PRF \citep[PRF-like photometry;][ Eq. 7]{Poleski_method}. Here we used both approaches and note that PRF was not elongated along the asteroid motion. We run the MCPM for every epoch and every asteroid separately. To find linear coefficients used to subtract instrumental trends we use all epochs except $\pm2~\mathrm{d}$ around the epoch of interest. This ensures that an asteroid does not affect the calculation of the linear coefficients. The pixels against which decorrelation is done are selected separately for each run, hence, they change as an asteroid moves on the sky. 
For each target pixel we assumed a model of astrophysical flux that is equal to 0 for all epochs except $\pm2~\mathrm{d}$ around the epoch of interest. \citet{Poleski_method} found that typical scatter of the flux is on the order of $20~\mathrm{e^-s^{-1}}$ (these are units in which \textit{K2} data products are distributed) and here we use this estimate. We also found that there is numerical noise at the level of $1~\mathrm{e^-s^{-1}}$, which is negligible compared to $20~\mathrm{e^-s^{-1}}$.

\section{Method of period determination and its uncertainty}

The photometry extracted with MCPM contains outlying points and long-term trends. The latter are most likely caused by the motion of the asteroid during an exposure. Thus, before determination of the rotation period we have removed the instrumental effects that may influence the results. In order to remove outliers, all points that differ from the local mean ($\pm 100$ points) by more than $3 \sigma$ were rejected. We have removed also the systematic trends based on the Simple Moving Average method where we calculate the mean flux of the surrounding 50 points and subtract it from each data point.
The magnitudes of measured points were corrected for the distance of the asteroid from both, the Kepler satellite and the Sun, so that an absolute magnitude is determined corresponding to a distance of 1 AU from both, the Sun and the \emph{Kepler} satellite. 


The rotation periods were determined using two independent codes. 
We have used discrete Fourier transform, which is a very efficient way to estimate the periodogram of unequally spaced data \citep{Kurtz85}. This method was implemented via the FNPEAKS code. For comparison, we also fit periodic orthogonal polynomials and apply analysis of variance \citep{Czerny2} as implemented in the AOV software. In this method for any number of observations we know the probability of the distribution, so this is very useful for small samples of data. 



\begin{figure}
\begin{center}
\includegraphics[width=\columnwidth]{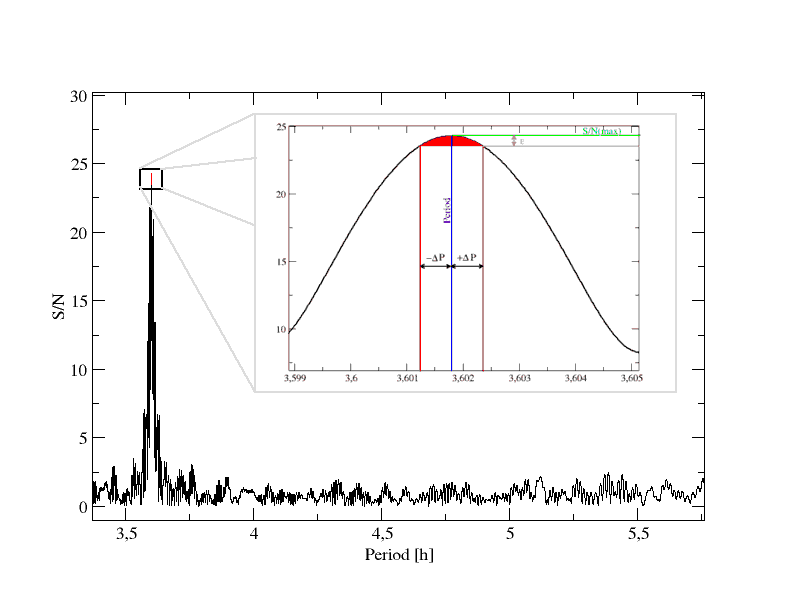}
\end{center}
\caption{\label{ddd}{Example periodogram illustrating the uncertainty calculation. }}
\end{figure}

As well as the rotation periods, we derive their uncertainty.
For the maximum value of the $S/N$
in the periodogram, we take the maximum height of the peak 
and find the mean noise power in the form \citep{Bartczak}:
\begin{equation}
   \epsilon=\frac{(S/N)_{max}}{\sqrt{N-n}}, 
\end{equation}
where $N$ denotes the number of observations, and $n$ denotes the number of parameters. Then, the $\epsilon$ was used to determine the confidence level and to calculate the uncertainty of period determination 
 \begin{equation}
(S/N)_{accepted}= (S/N)_{max} - \epsilon.
\end{equation}
The width of the line is $1 \sigma$ confidence interval of the oscillation period.
The graph presenting procedure of uncertainty determination is shown in Fig \ref{ddd}. It was described in details in \cite{Czerny1991}.

\section{Results}
\begin{figure}
\begin{center}
\includegraphics[width=0.49\columnwidth]{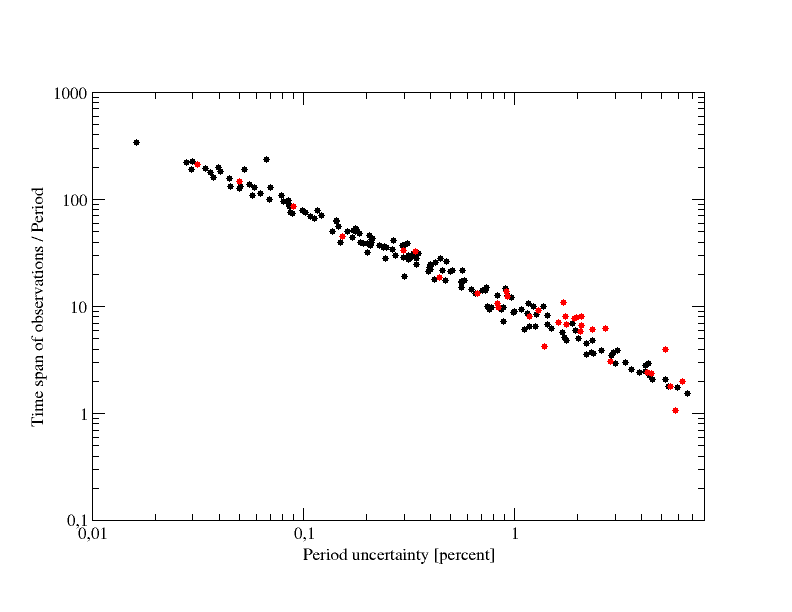}
\includegraphics[width=0.49\columnwidth]{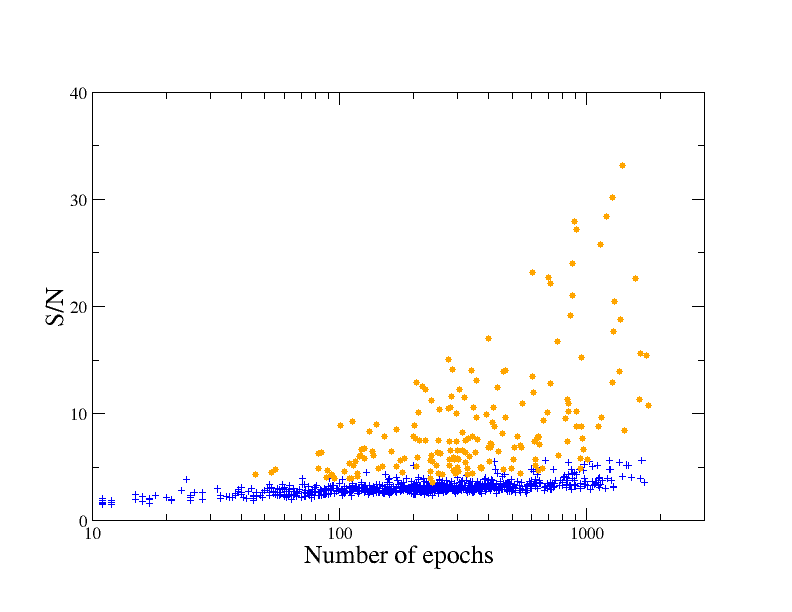}
\end{center}
\caption{\label{fig1}{Left: The rotation period uncertainty vs. ratio of the time span of the observation to the calculated rotation period. Black dots denote the objects with certain rotation period determination, red dots the objects with less certain estimation. Right: Signal to noise ratio as a function of number of photometric points. Yellow dots show objects for which the aperture-like photometry and PRF-like photometry gave consistent results. Blue dots mark other objects.}}
\end{figure}

To determine rotation periods for a large sample of asteroids
we have investigated 1026 objects  
passing in the \textit{Kepler}'s field of view during the \textit{K2}C9.
Using the procedure described in previous Section we have calculated the $S/N$ ratio, and tried to find the best-fitting period. 
The number of data points for some of the asteroids is too small to allow reliable period measurement. 
We have 205 asteroids with fewer than 100 measured points after removing outliers. Among them we have taken into account only those displaying large amplitude of brightness changes, but they still should be considered as uncertain. We have included only the objects for which the derived rotation period is a few times shorter than the time span of observations. In left panel of Fig \ref{fig1} we show the uncertainty of period determination as a function of the ratio of  the observational timespan to the rotation period. 
The Figure clearly shows that the uncertainty drops significantly if the rotation period is covered many times during the observational time span.

For the asteroids with low $S/N$ (here $S/N$ is used in context of the data) the determination of an unique
rotation period was not possible. As we mentioned, these objects spent too little time in the \textit{Kepler}'s field of view, which resulted in too 
small a number of measured points. Moreover, some objects
displayed very small variations in brightness, thus the signal
 was at a level similar to the noise, so
for these asteroids determination of a unique
rotation period was also not possible. 
Such small amplitude may indicate that the shape of the body is close to spherical, 
or we are looking at one of the poles of the asteroid.
The right panel of Fig. \ref{fig1} shows the $S/N$ as a function of number of measured points. Yellow dots show objects for which the aperture-like photometry and the PRF-like photometry gave consistent results. Blue dots mark objects for which the results were different. It can be seen that the high signal to noise ratio is easily achieved for objects with a large number of photometric points.   

As a trustworthy we adopt the results for which S/N ratio is relatively high ($>5$) or number of points is high enough to cover rotation period many times. The objects fulfilling these criteria which give consistent results with both codes and for both types of photometry are considered as certain.
In order to prove that our methods and results are reliable, we have compared the
rotation periods determined from the \textit{K2}C9 data with values found in the literature. For this purpose, we queried  the Asteroid Lightcurve Database \citep[LCDB;][]{Warner}. 
In the notation adopted in the LCDB, the quality code 1 means that the determination is very poor, and 3 denotes that the rotation period is certain.
In Table \ref{tab:lcdb} we show the values from the LCDB together with the quality code informing about the reliability of the determined period,
and compare them with our results.  In most cases the agreement is very good within adopted uncertainty.
The only cases where our results disagree are 2088, 2662, 4993, 87630, 97811 and 169114. For  asteroid 2662 the rotation period in the LCDB has quality code 1+, which means that rotation period is very uncertain and it is likely wrong.  The period of $39.68$ ~h, that we have found is based on a large number of data points (1001), and shows twice the value given in the LCDB. In this case, the $S/N$ value is not very high (5.7) with rather small amplitude of brightness changes, but both photometry extractions resulted in the same value of period, which was confirmed with both codes.
A similar situation exists for 97811, where we have found a resonant frequency with the period ratio of our calculation to the LCDB value as 3:4. In this case, the $S/N$ is quite high and the lightcurve with two minima and two maxima is clearly visible in the fitted period.
We cannot say too much about the difference in obtained periods for 2088, 4993, 87630 and 169114. According to the adopted criteria, 2088 is classified as probable but not certain in our Table \ref{tab:results2}.  
Asteroid 4993 is marked with $2-$ quality code in LCDB. In our data number of measured points is very low (85), but it always 
gives good agreement between photometry extracted using both methods obtained with both codes, pointing to a rotation period of $7.3$ ~h, so for this object both the period in the LCDB and the one calculated here are rather uncertain.
In contrary to the asteroid mentioned above, the high number of points (1395) and clear sinusoidal trend in the lightcurve of 87630 makes the result trustworthy for this asteroid. The similar situation exists for 169114 although here the number of points is lower (412).

\startlongtable
\begin{deluxetable*}{ccccc}
\tablecaption{Comparison of asteroids rotation period from the LCDB and determined in this paper. The first column shows asteroid number, the second and third show rotation period from the LCDB with quality code, the fourth shows period determined in this paper together with uncertainties, and the fifth shows the $S/N$ ratio.\label{tab:lcdb}}
\tablewidth{0pt}
\tablehead{
\colhead{Asteroid} & \colhead{LCDB Period} & \colhead{LCDB quality} & \colhead{Period} & \colhead{$S/N$} \\
\colhead{number} & \colhead{(h)} & \colhead{flag} & \colhead{(h)} & \colhead{}
}
\decimalcolnumbers
\startdata
(271) Penthesilea & 18.787 & 3 & $18.58\pm{0.55}$ & 5.8 \\ 
(1804) Chebotarev & 4.026 & 3 & $4.0241\pm{0.0025}$ & 24.0 \\ 
(1920) Sarmiento & 4.048 & 3 & $4.043\pm{0.011}$ & 6.5 \\ 
(2088) Sahlia & 10.37 & 2 & $58.3\pm{2.6}$ & 5.5 \\ 
(2662) Kandinsky & 20.43 & $1+$ & $39.68\pm{0.27}$ & 5.7 \\ 
(3751) Kiang & 8.2421 & $3-$ & $8.212\pm{0.058}$ & 6.6 \\ 
(3989) Odin & 5.3229 & 3 & $5.348\pm{0.069}$ & 4.3 \\ 
(4631) Yabu & 7.356 & 3 & $7.354\pm{0.054}$ & 12.9 \\ 
(4993) Cossard & 48.26 & $2-$ & $7.27\pm{0.15}$ & 6.4 \\ 
(5026) Martes & 4.4243 & 3 & $4.425\pm{0.012}$ & 12.5 \\ 
(5114) Yezo & 4.3 & 2 & $4.3326\pm{0.0025}$ & 15.2 \\ 
(6825) Irvine & 3.61588 & 3 & $3.6148\pm{0.0064}$ & 14.1 \\ 
(9213) 1995 UX5 & 36.2963 & 2 & $36.19\pm{0.79}$ & 11.6 \\ 
(9582) 1990 EL7 & 5.79899 & 2 & $5.802\pm{0.020}$ & 10.5 \\ 
(9780) Bandersnatch & 7.06329 & $3-$ & $7.052\pm{0.040}$ & 5.8 \\ 
(10125) Stenkyrka & 5.597 & 2 & $5.600\pm{0.020}$ & 10.4 \\ 
(10688) Haghighipour & 9.4 & 2 & $9.042\pm{0.044}$ & 7.5 \\ 
(10817) 1993 FR44 & 6.322 & 2 & $6.3230\pm{0.0019}$ & 30.2 \\ 
(13229) Echion & 8.46 & $2+$ & $8.4509\pm{0.0038}$ & 13.9 \\ 
(17966) 1999 JS43 & 78.1156 & 2 & $78.10\pm{0.49}$ & 23.2 \\ 
(18418) Ujibe & 3.47 & 2 & $3.473\pm{0.010}$ & 3.9 \\ 
(20694) 1999 VT82 & 3.322 & 2 & $3.3250\pm{0.0048}$ & 5.1 \\ 
(23031) 1999 XX7 & 3.075 & 3 & $3.059\pm{0.052}$ & 4.5 \\ 
(35815) 1999 JO48 & 9.74307 & 2 & $9.67\pm{0.52}$ & 4.3 \\ 
(36700) 2000 RT17 & 96.6775 & 2 & $94.8\pm{2.9}$ & 10.5 \\ 
(37393) 2001 XF24 & 10.04107 & 2 & $10.10\pm{0.18}$ & 6.3 \\ 
(44957) 1999 VG78 & 5.828 & 2 & $5.820\pm{0.023}$ & 6.3 \\ 
(50971) 2000 GP88 & 3.738 & 2 & $3.7394\pm{0.0078}$ & 8.3 \\ 
(77590) 2001 KM17 & 5.892 & 2 & $5.904\pm{0.055}$ & 6.8 \\ 
(84229) 2002 SH15 & 32.5815 & 2 & $32.57\pm{0.21}$ & 14.0 \\ 
(87630) 2000 RE55 & 34.4 & 2 & $10.1546\pm{0.0088}$ & 33.1 \\ 
(88472) 2001 QW111 & 5.476 & 2 & $5.472\pm{0.024}$ & 8.9 \\ 
(89347) 2001 VS66 & 9.092 & 2 & $9.09\pm{0.21}$ & 4.3 \\ 
(95666) 2002 GZ141 & 4.016 & 2 & $4.0154\pm{0.0016}$ & 11.3 \\ 
(97674) 2000 FJ50 & 8.161 & 2 & $8.169\pm{0.033}$ & 7.9 \\ 
(97811) 2000 OH33 & 44.0173 & 2 & $33.70\pm{0.42}$ & 7.5 \\ 
(109118) 2001 QH42 & 7.898 & 2 & $7.87\pm{0.14}$ & 3.9 \\ 
(132598) 2002 JG142 & 10.237 & 2 & $10.34\pm{0.20}$ & 4.4 \\ 
(169114) 2001 OK47 & 6.86 & 2 & $7.880\pm{0.015}$ & 7.1 \\ 
(271713) 2004 RU200 & 4.19 & 2 & $4.1974\pm{0.0028}$ & 10.1 \\ 
(382243) 2012 SV39 & 7.2 & $1+$ & $7.2166\pm{0.0033}$ & 9.6 \\ 

\enddata
\end{deluxetable*}

Among all investigated asteroids we have determined the synodic rotation periods (from vantage point of \textit{Kepler}) for 188 asteroids, and for most of these objects $P$ is determined for the first time. The 155 objects for which we have found the most trustworthy rotation periods are listed in Table \ref{tab:results1}. These are objects with a high $S/N$ ratio where both types of photometry (aperture-like and PRF-like) and both numerical codes (FNPEAKS and AOV) gave the same results within their mutual adopted uncertainties. For these objects, we usually have 
a large number of photometric points, and the lightcurves display large amplitudes. In Table~\ref{tab:results2} we include 33 objects for which
 the $S/N$ ratio is lower, but all results obtained with both codes are consistent, or 
 at least one photometric dataset gave results (confirmed with both methods) with a reasonable $S/N$ (around 5), and with clearly visible shape of the lightcurve. 
In this group we have objects for which the rotation periods were also previously determined, so these are the test cases that our level of acceptance is reasonable. As mentioned above all objects for which we have determined the rotation periods from the \textit{K2}C9 data are listed in Tables \ref{tab:results1} and \ref{tab:results2}. We provide there the number and the name of asteroid, the number of data points that were for our disposal, the length of the observations (in days), and the determined rotation period with uncertainties and $S/N$ ratio. All the results from our studies will be available in CDS (Strasbourg astronomical Data Center) after acceptance of the paper. We will also include there time series photometry for rejected objects for which the determination of rotation periods was not possible from the dataset that we have.

\begin{figure*}
\begin{center}
\includegraphics[width=0.49\columnwidth]{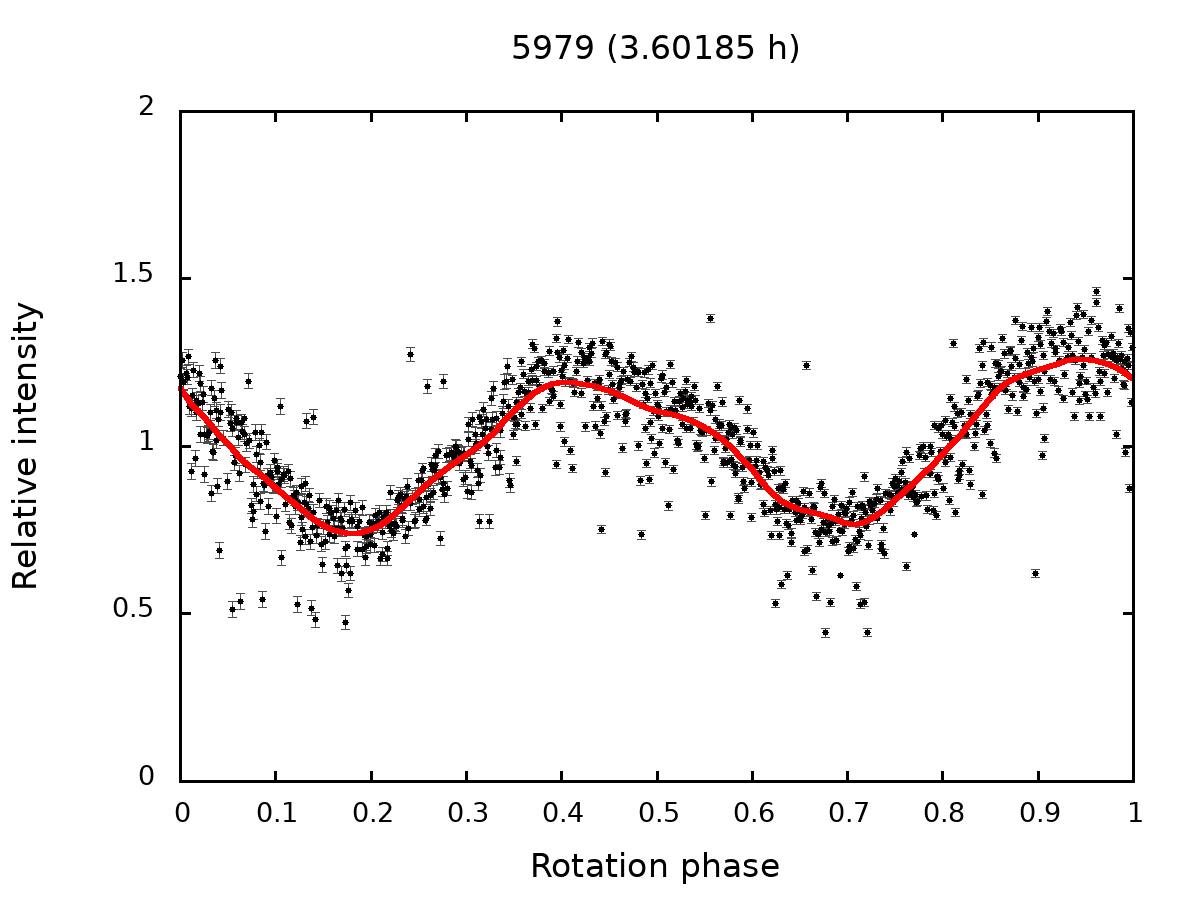}
\includegraphics[width=0.49\columnwidth]{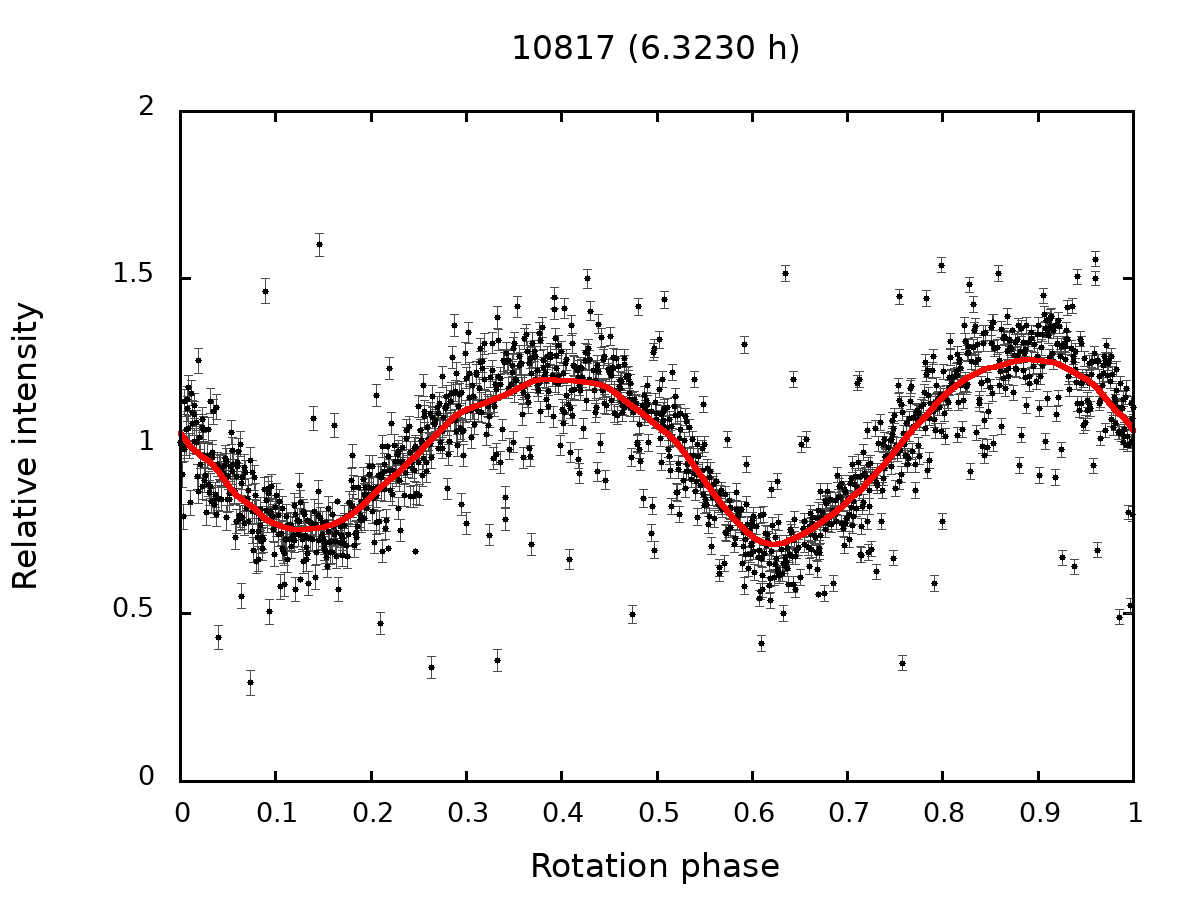}
\includegraphics[width=0.49\columnwidth]{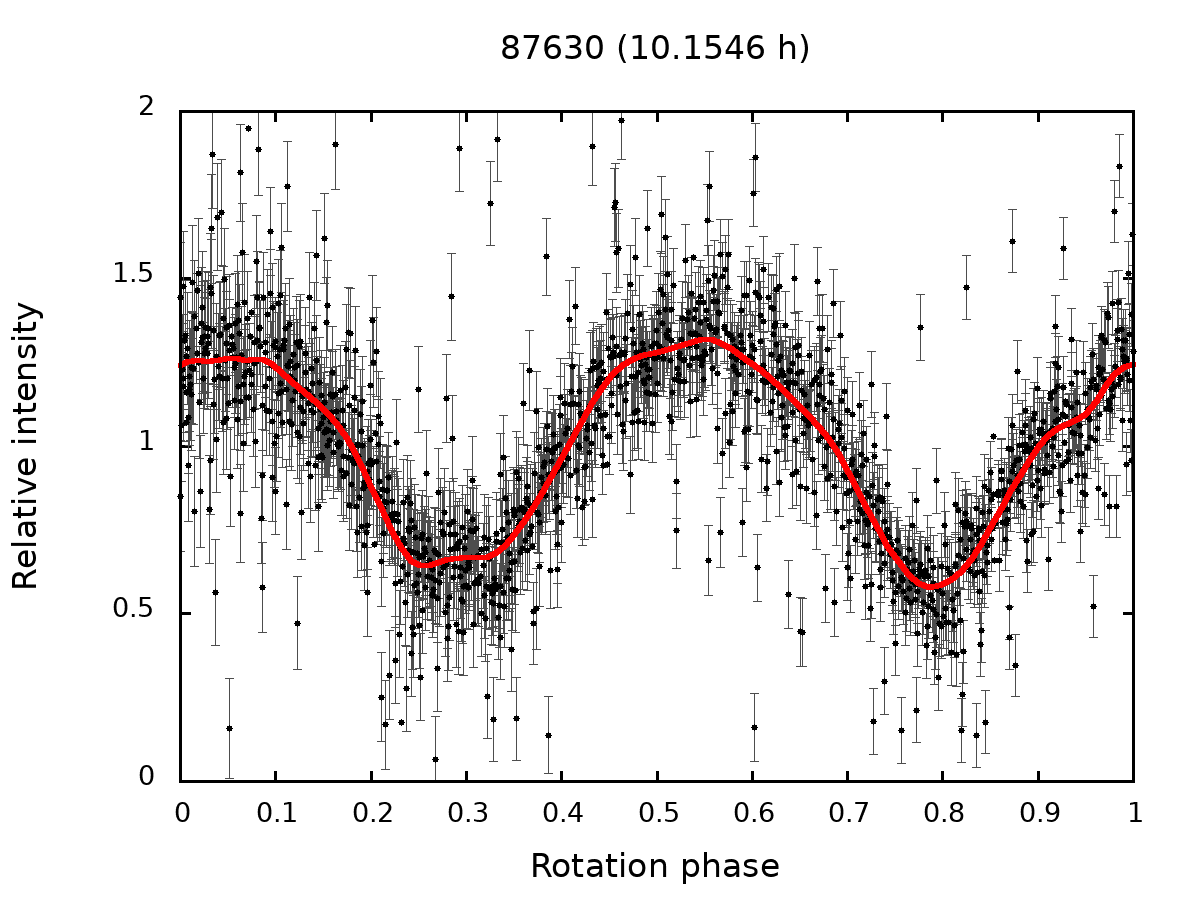}
\includegraphics[width=0.49\columnwidth]{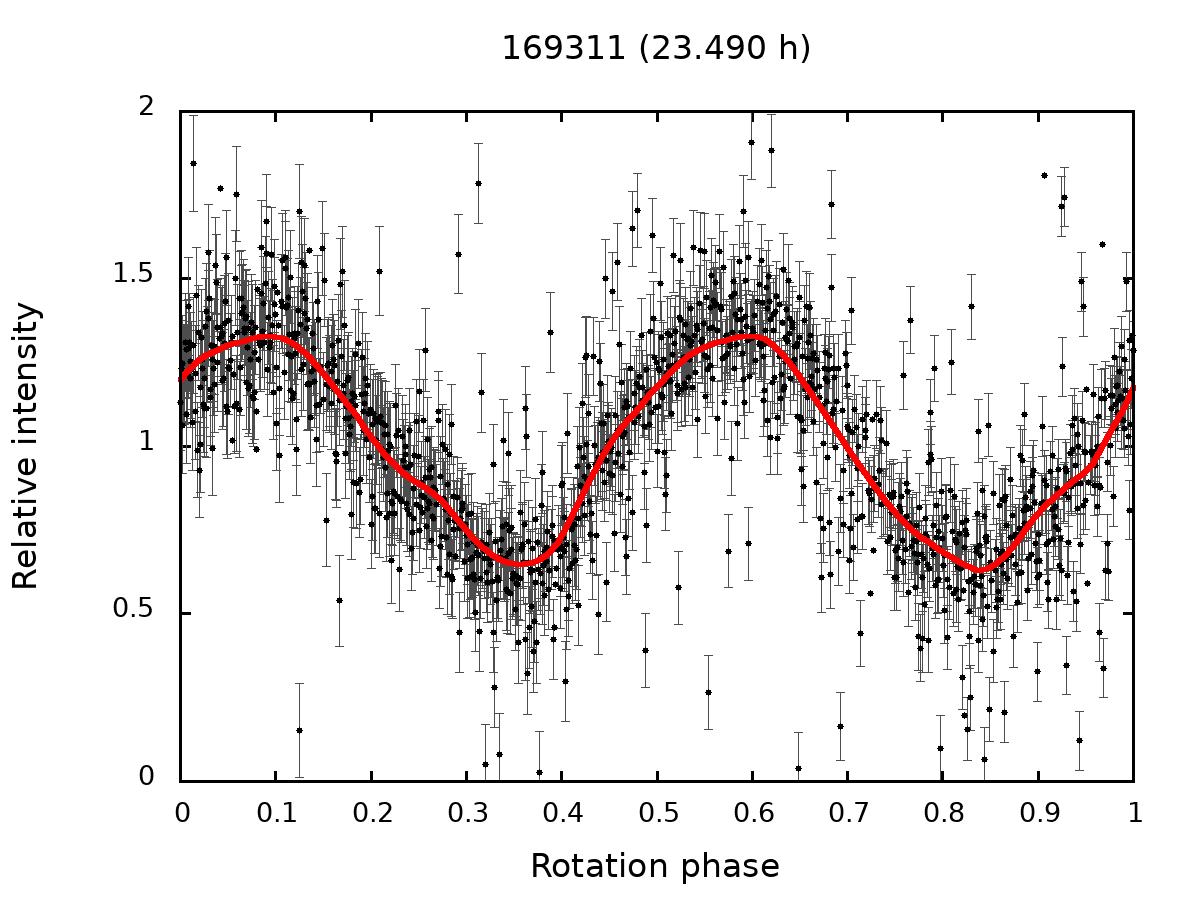}
\end{center}
\caption{\label{fig2}{The examples of lightcurves with the best signal to noise ratio ($S/N>20$) for asteroids 5979 (top left), 10817 (top right), 87630 (bottom left) and 169311 (bottom right). The red lines denote mean trend in the lightcurve. }}
\end{figure*}

\begin{figure*}
\begin{center}
\includegraphics[width=0.49\columnwidth]{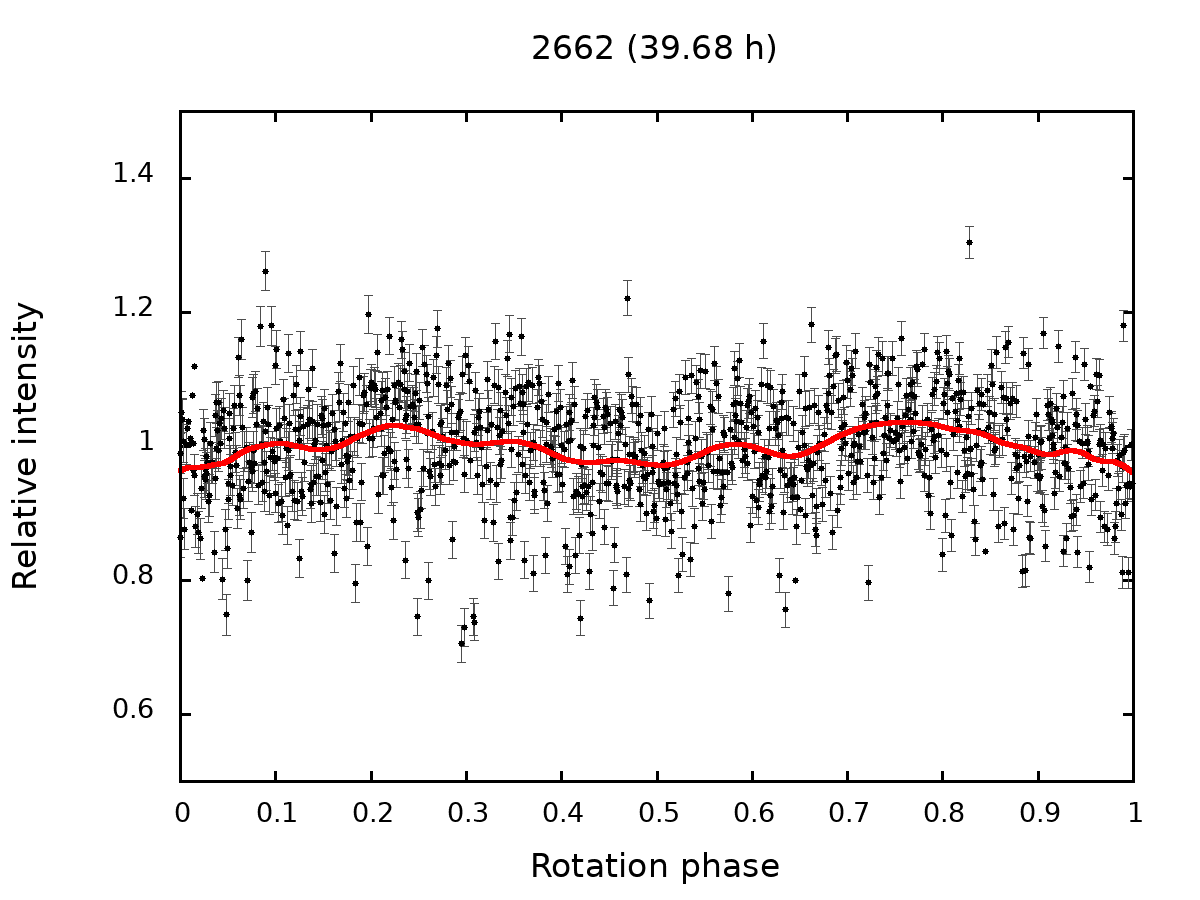}
\includegraphics[width=0.49\columnwidth]{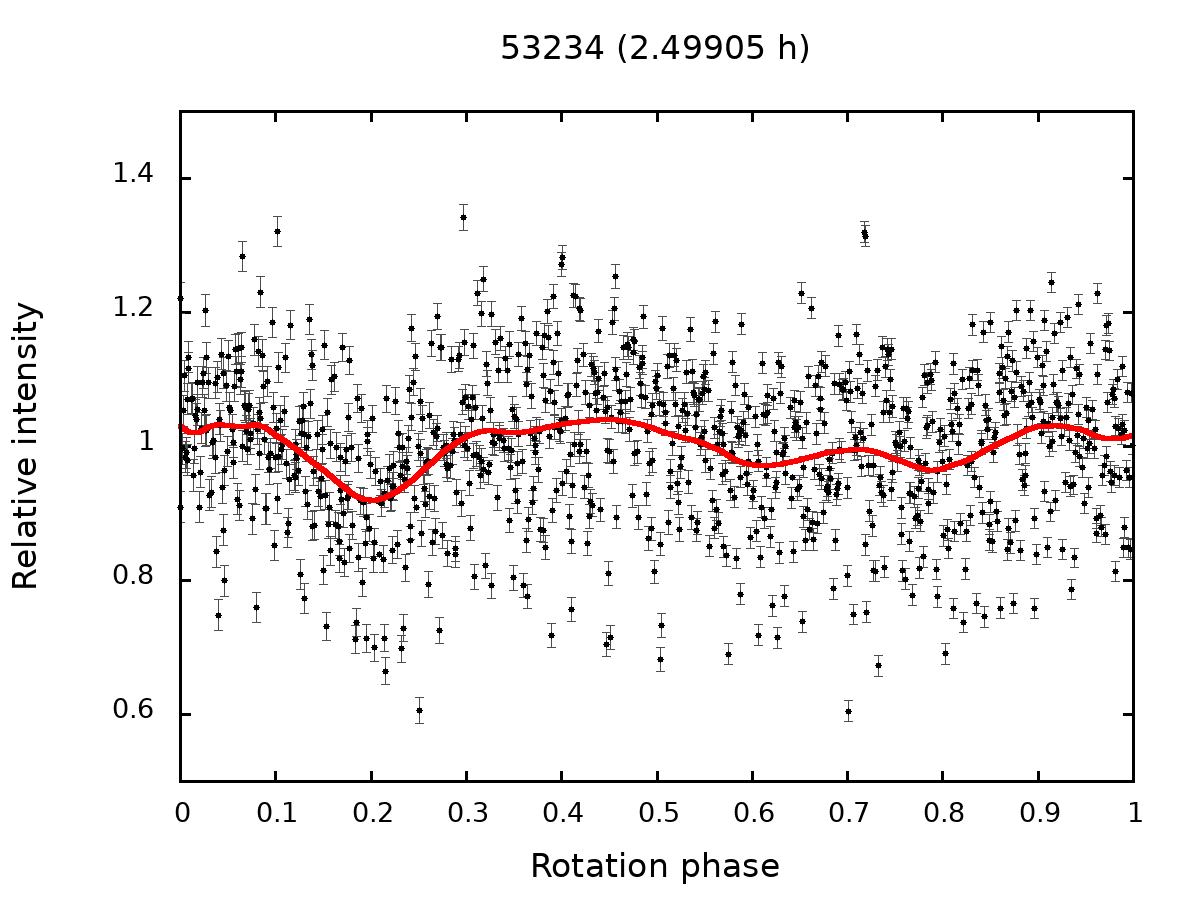}
\includegraphics[width=0.49\columnwidth]{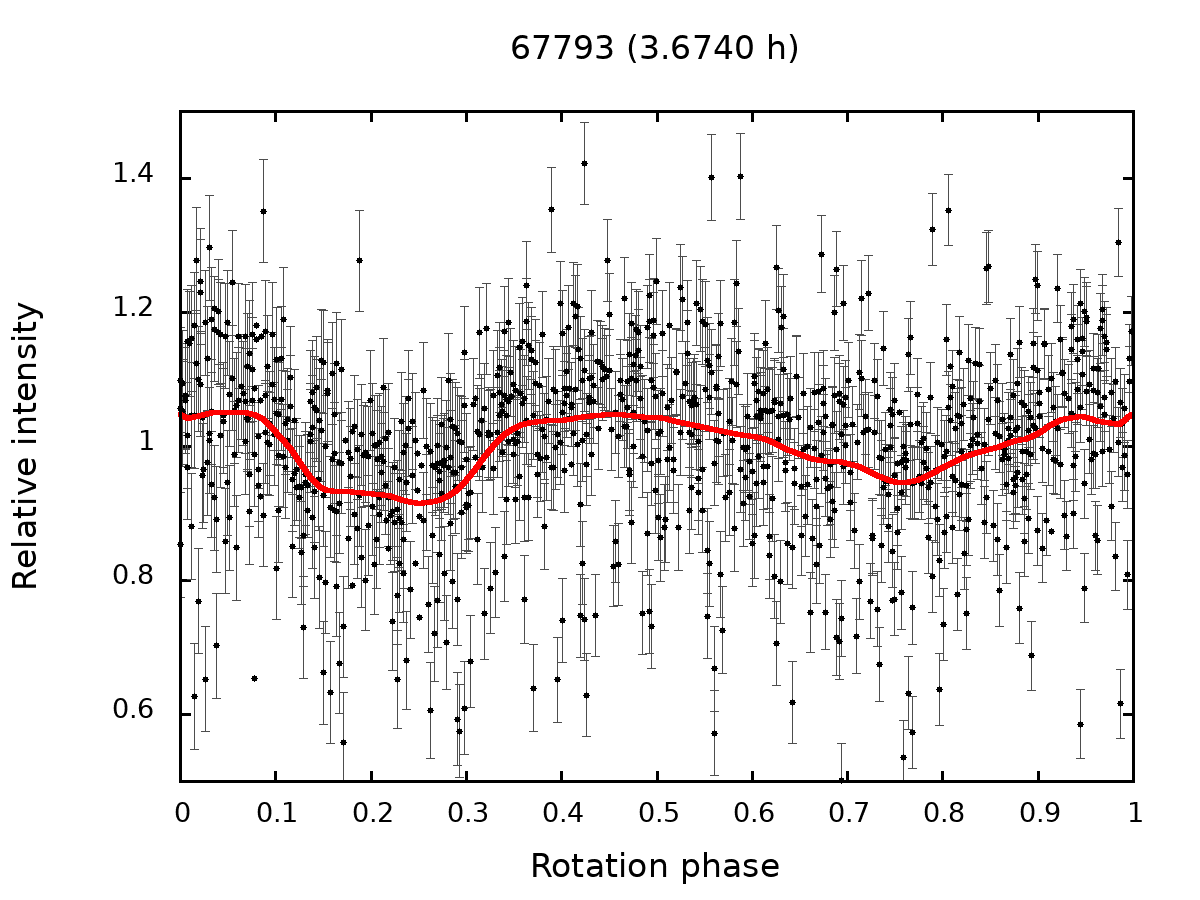}
\includegraphics[width=0.49\columnwidth]{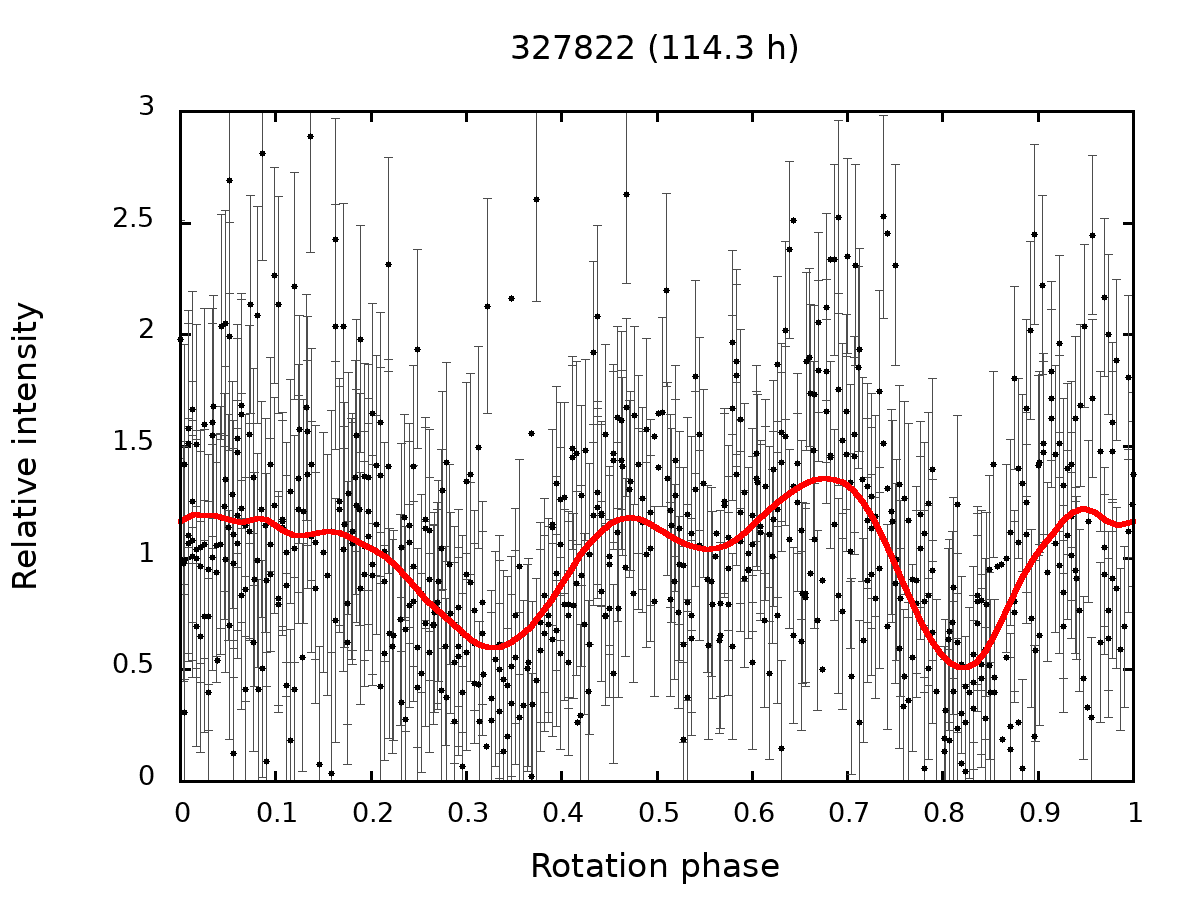}
\end{center}
\caption{\label{fig3}{The examples of lightcurves from Table \ref{tab:results2} with slightly worse signal to noise ratio ($S/N>20$) for asteroids 2662 (top left), 53234 (top right), 67793 (bottom left) and 327822 (bottom right). The red lines denote mean trend in the lightcurve.}}
\end{figure*}

Figure \ref{fig2} shows the examples of lighturves for objects with the highest $S/N$ ratio for which we expect very good period determination. Figure~\ref{fig3} shows for comparison the lightcurves for which estimated periods are less certain. The red lines in these Figures denote mean trend in the lightcurve. These objects are listed in Tables \ref{tab:results1} and \ref{tab:results2}, respectively.

\begin{figure*}
\begin{center}
\includegraphics[width=0.327\columnwidth]{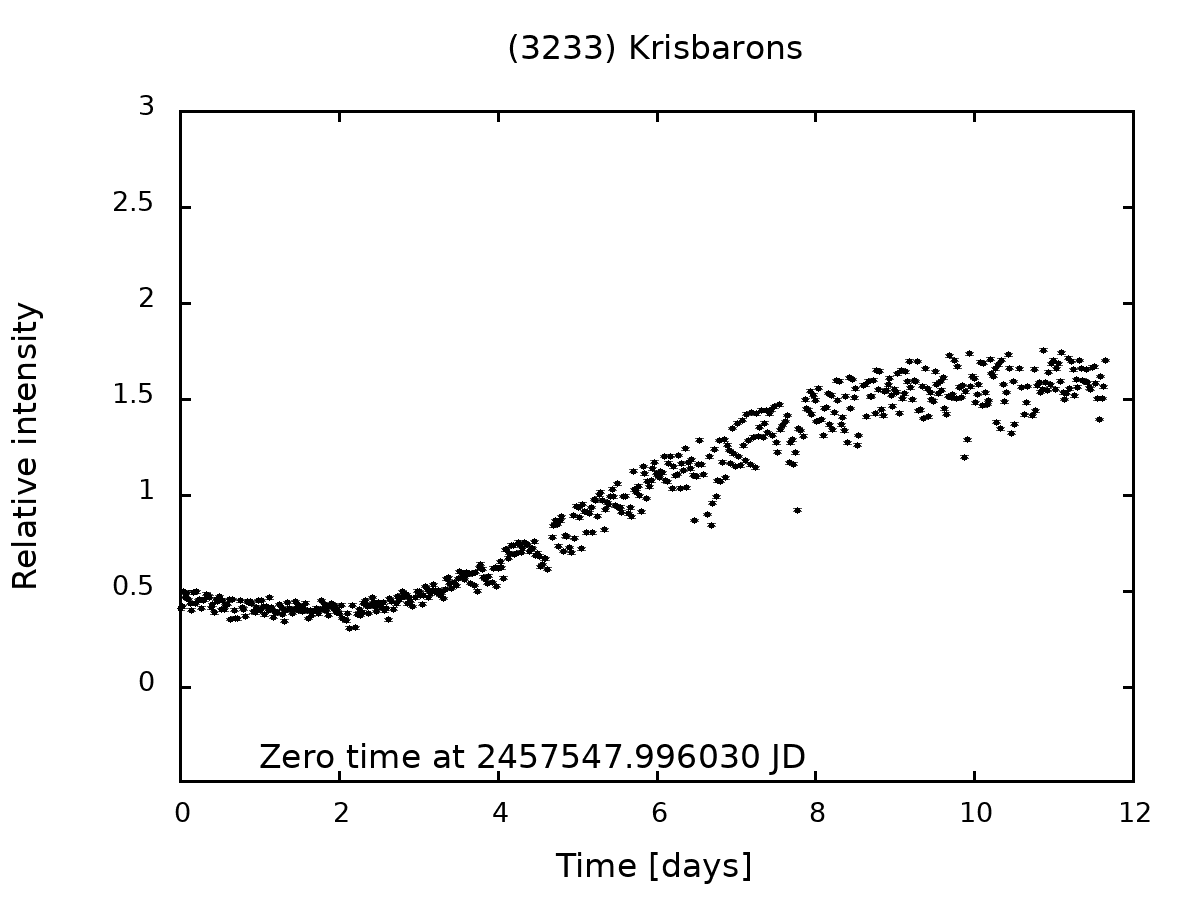}
\includegraphics[width=0.327\columnwidth]{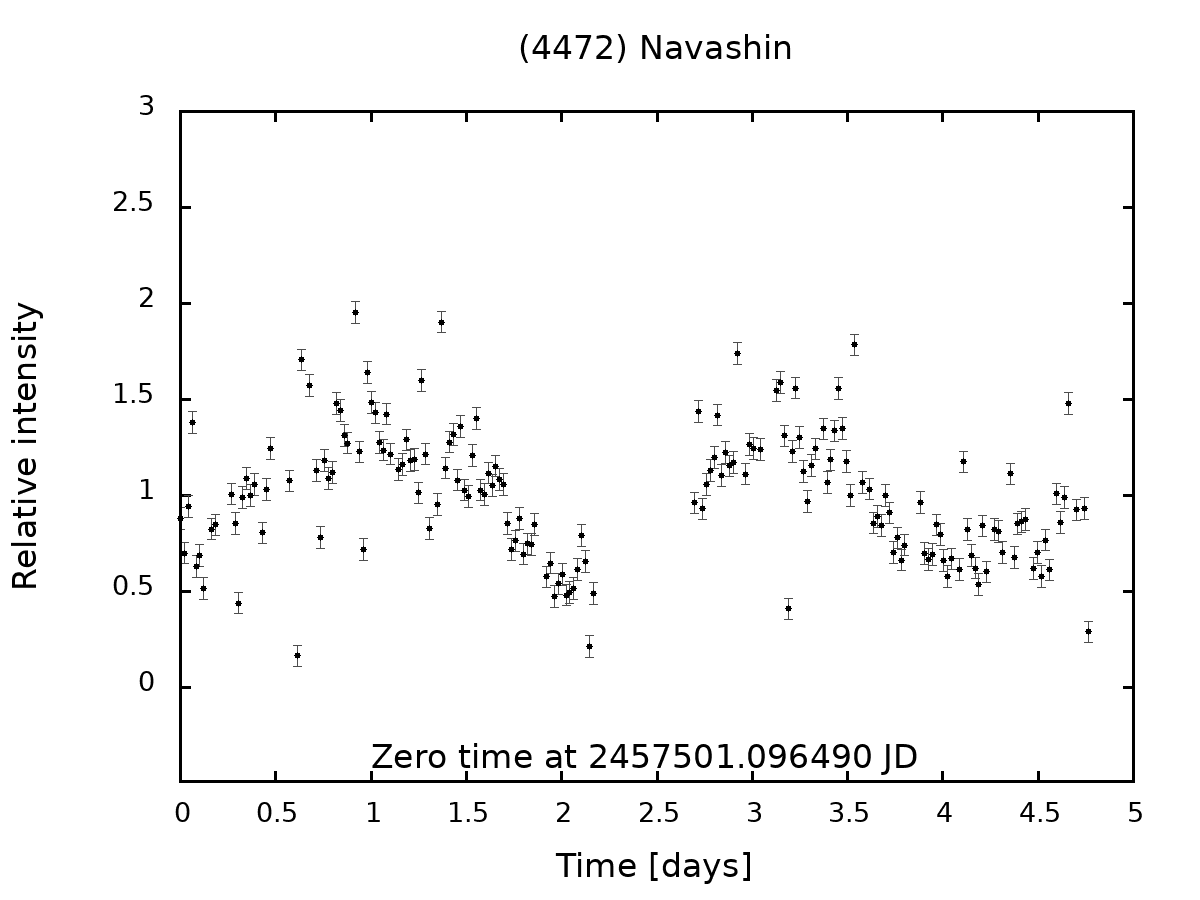}
\includegraphics[width=0.327\columnwidth]{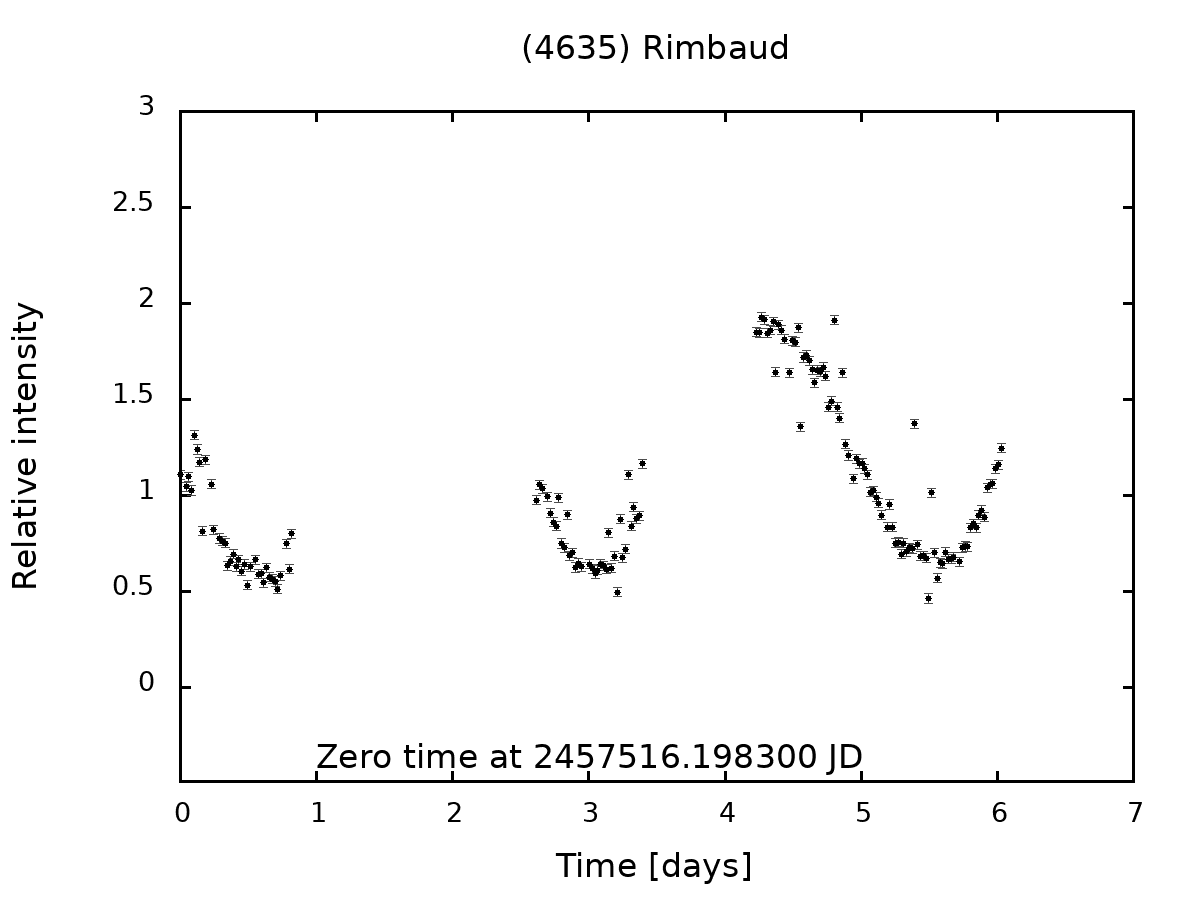}
\includegraphics[width=0.327\columnwidth]{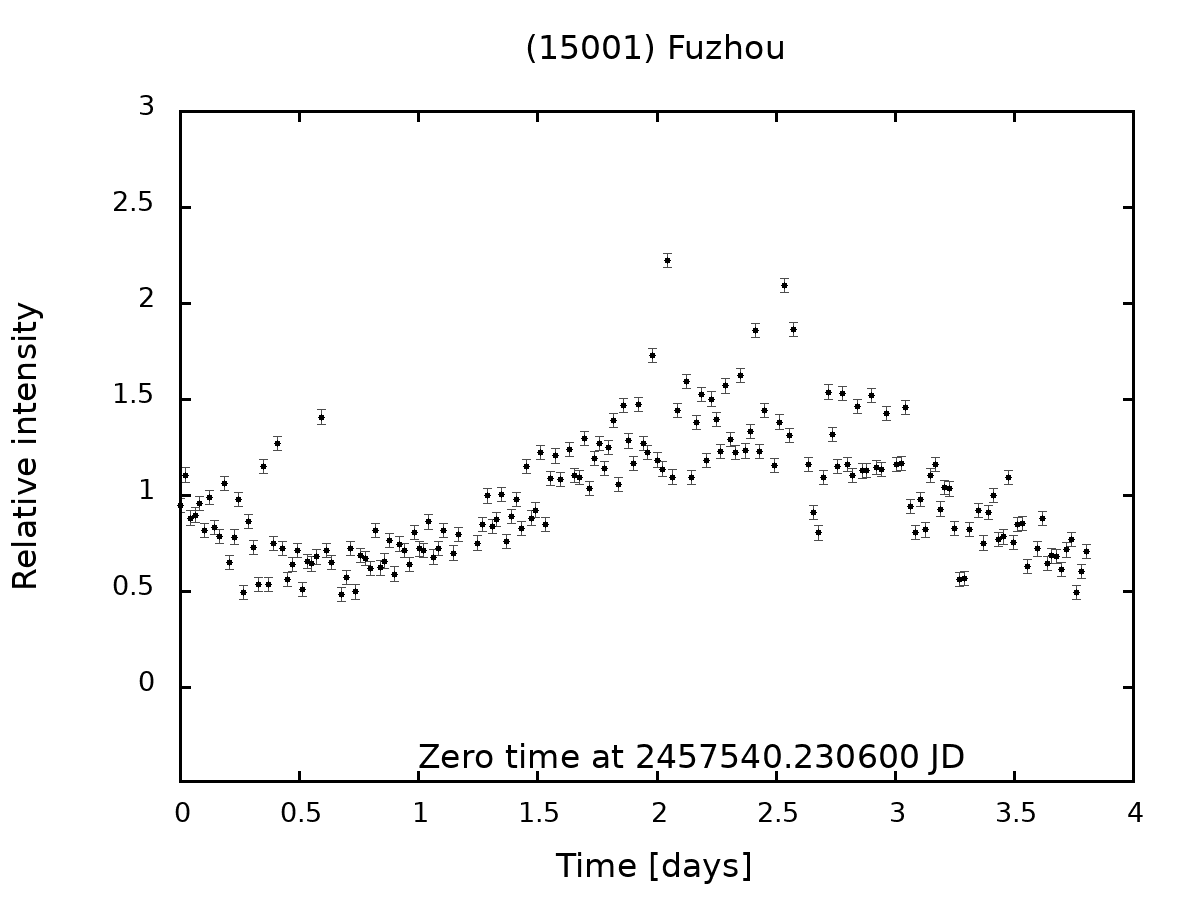}
\includegraphics[width=0.327\columnwidth]{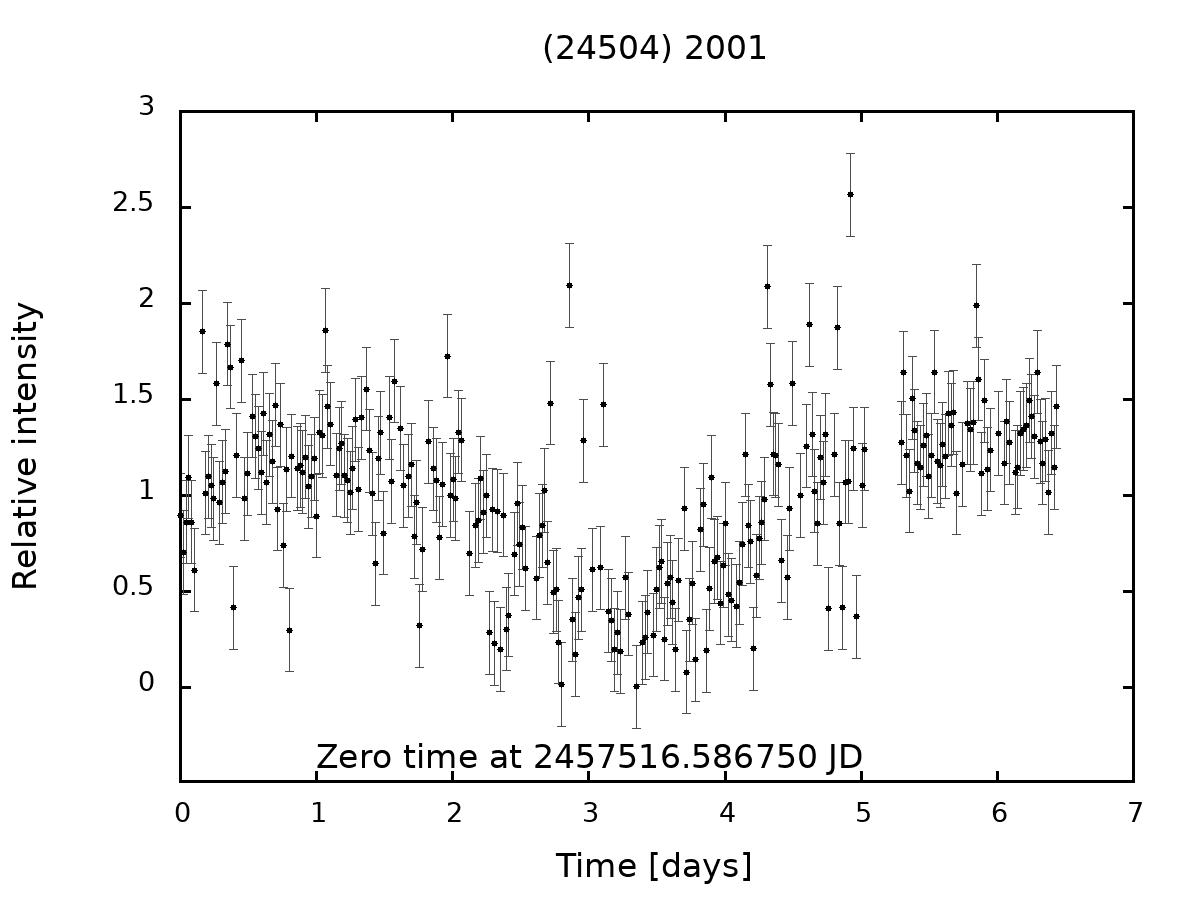}
\includegraphics[width=0.327\columnwidth]{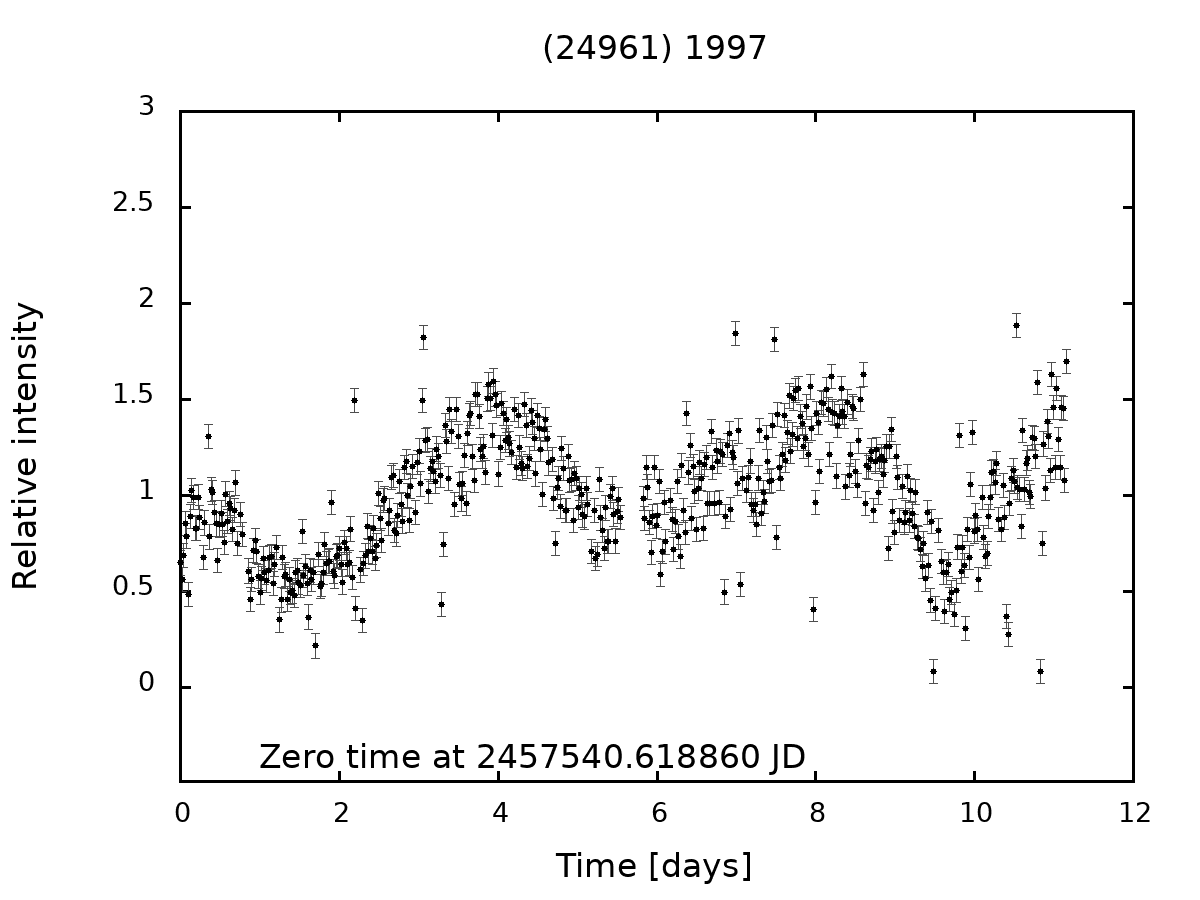}
\includegraphics[width=0.327\columnwidth]{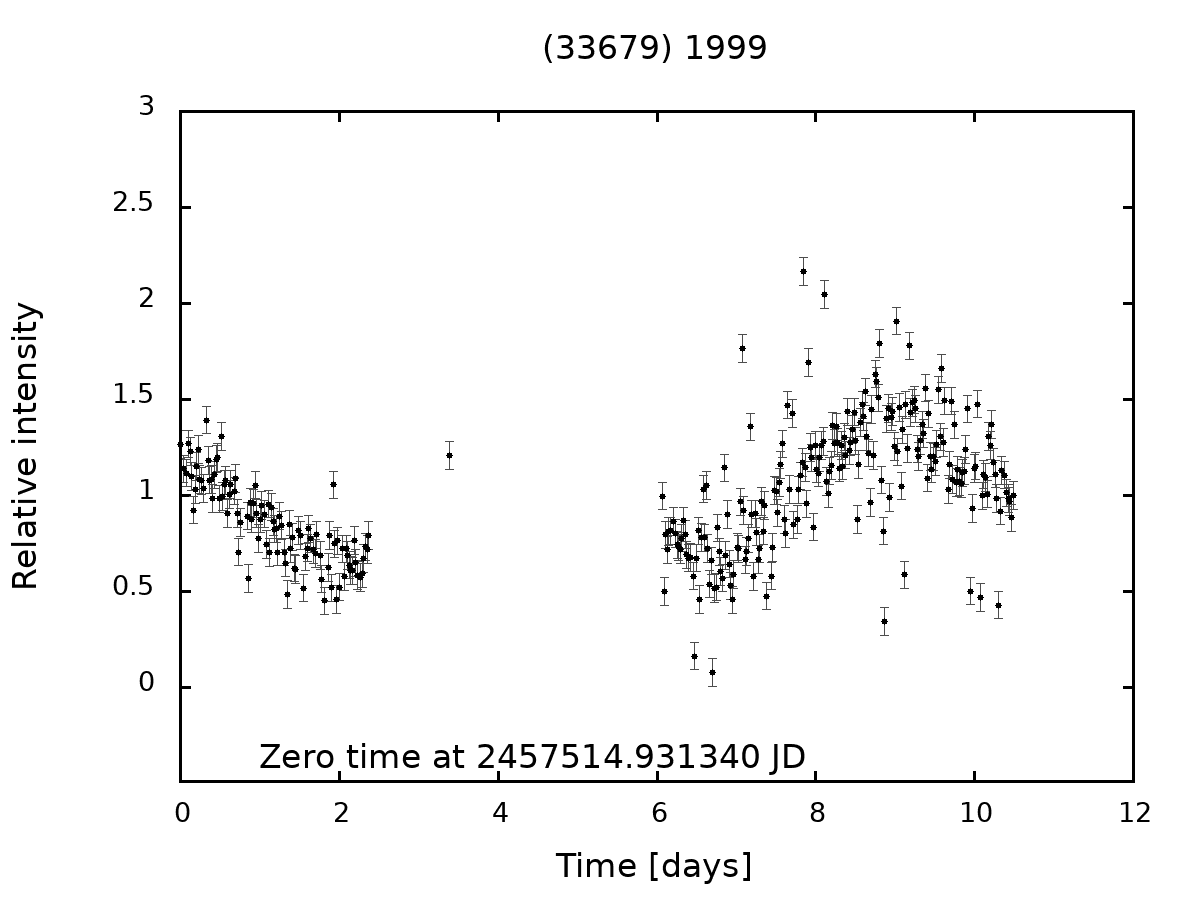}
\includegraphics[width=0.327\columnwidth]{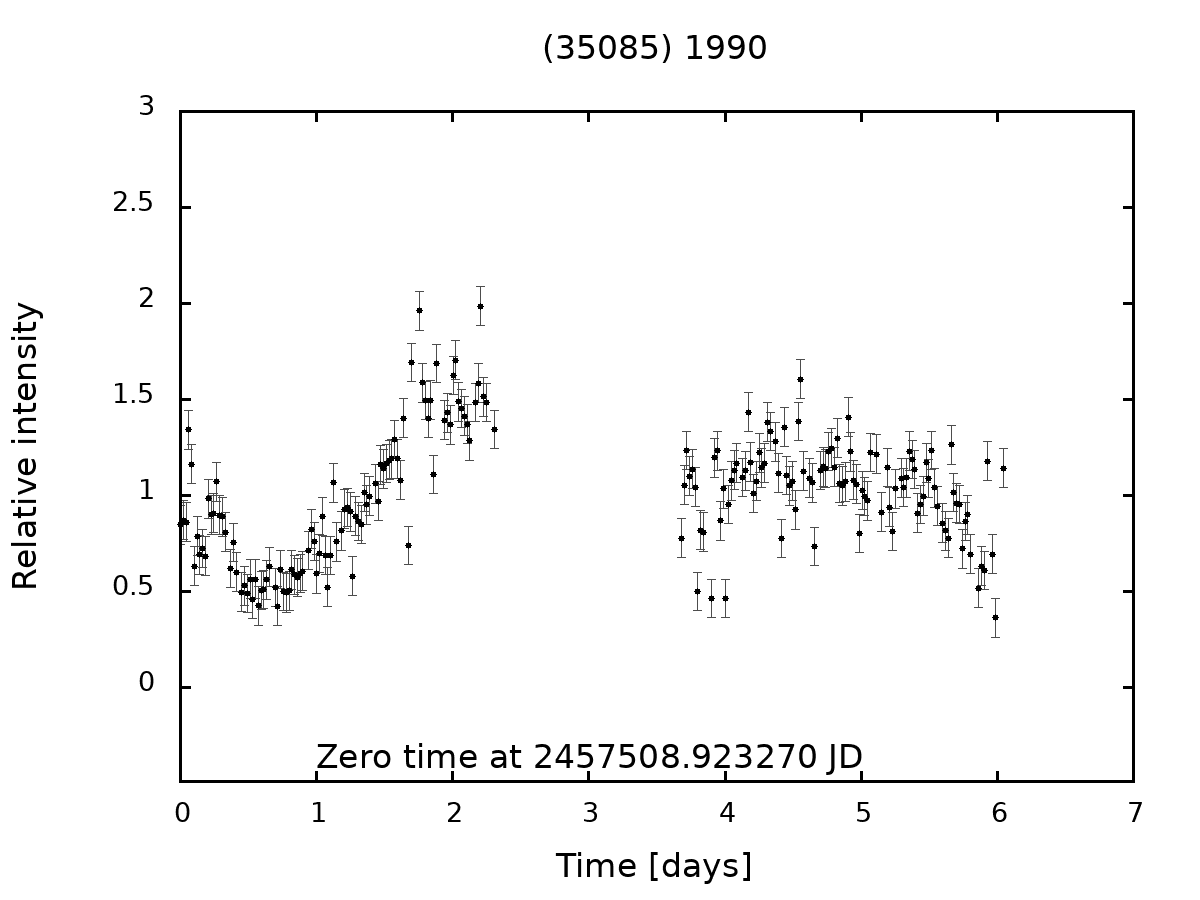}
\includegraphics[width=0.327\columnwidth]{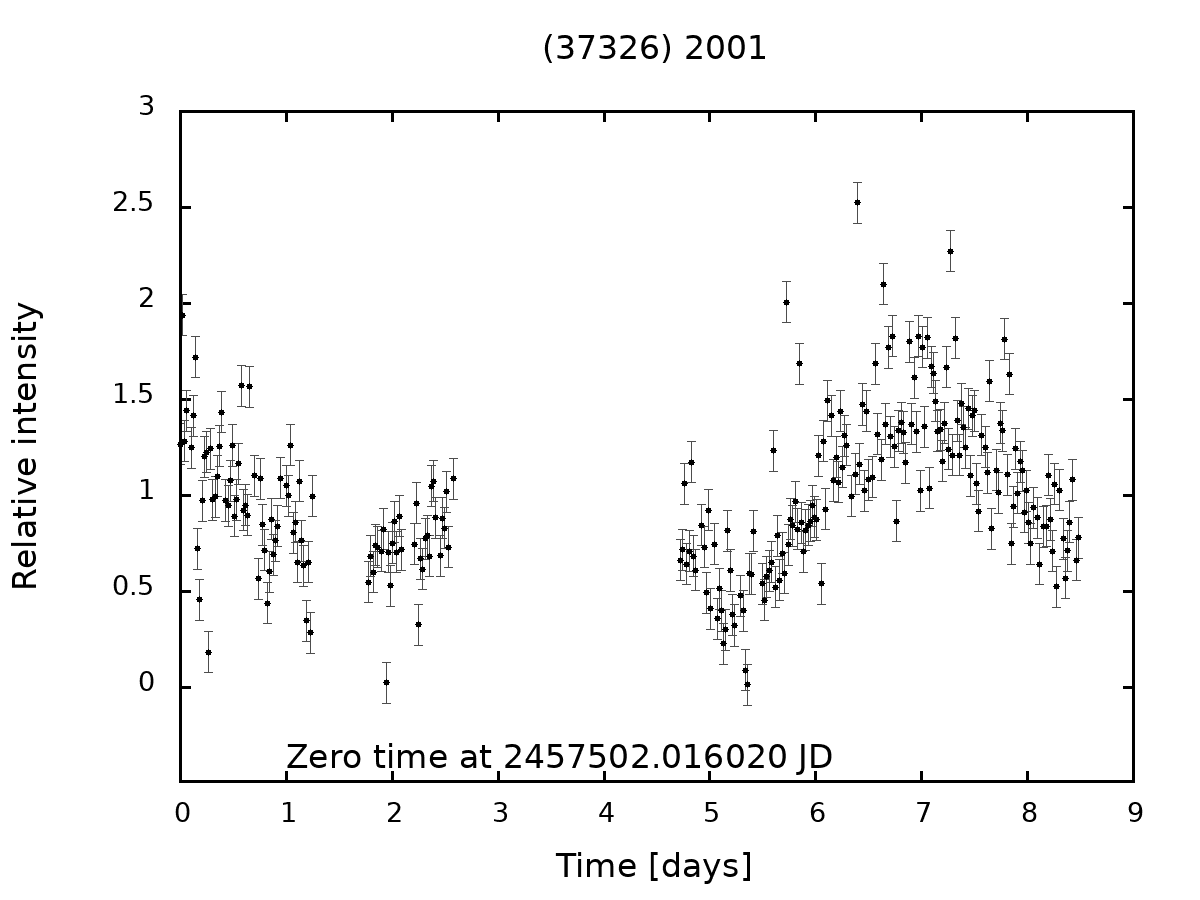}
\includegraphics[width=0.327\columnwidth]{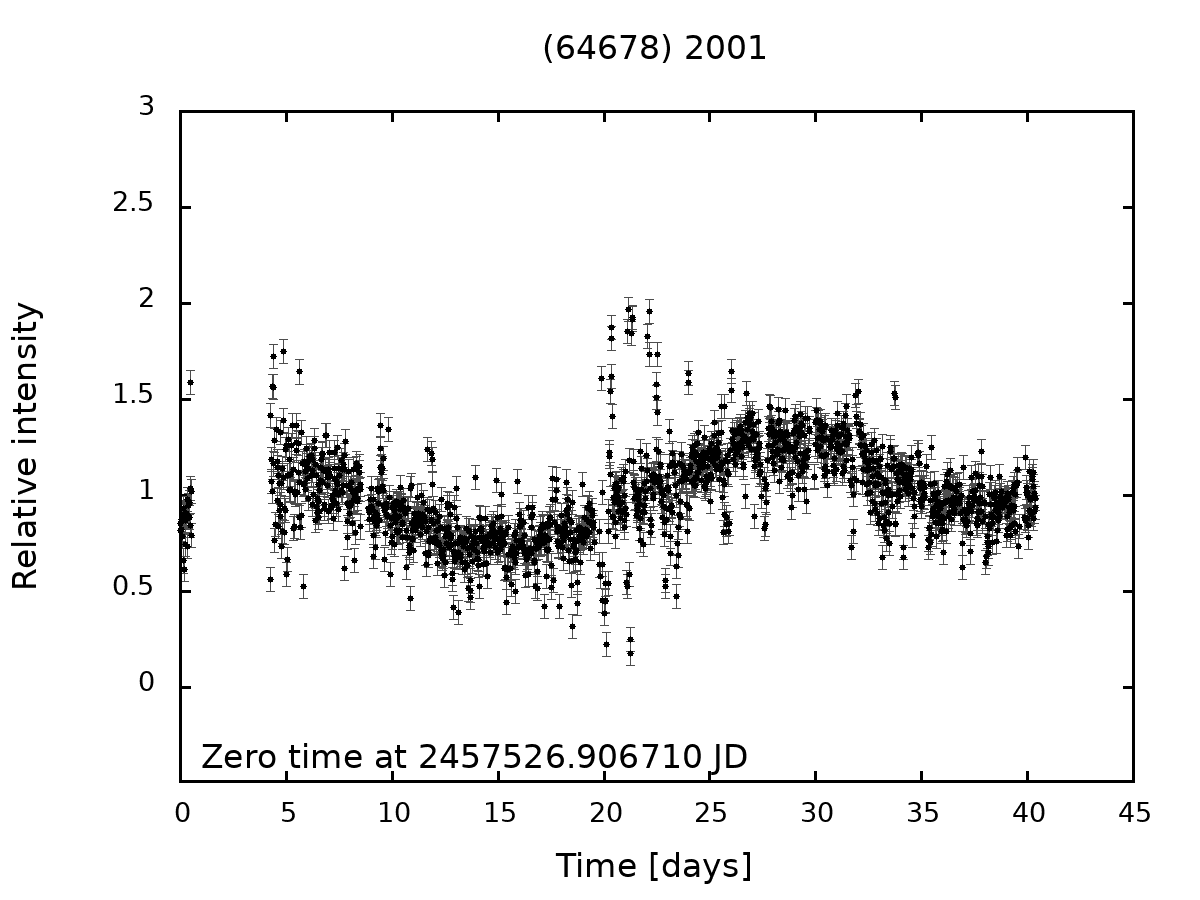}
\includegraphics[width=0.327\columnwidth]{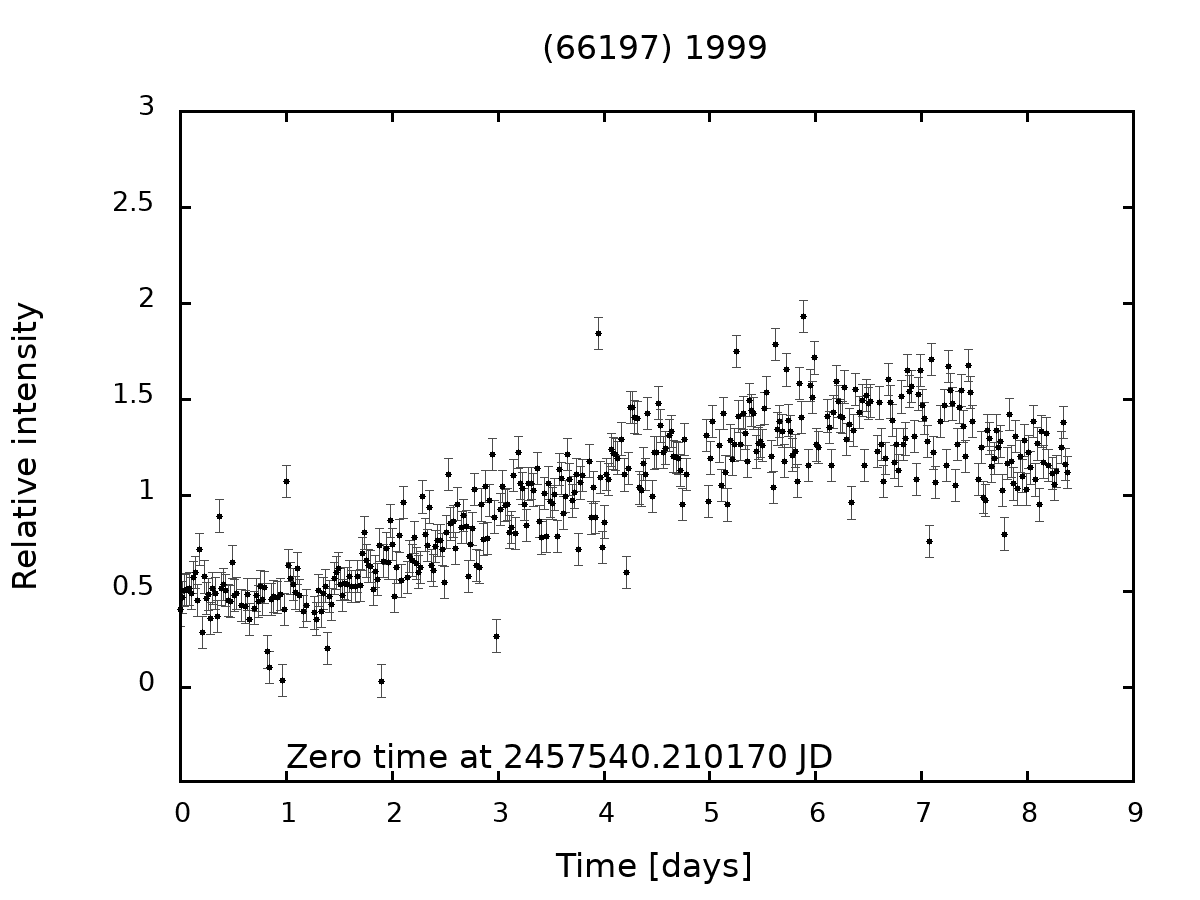}
\includegraphics[width=0.327\columnwidth]{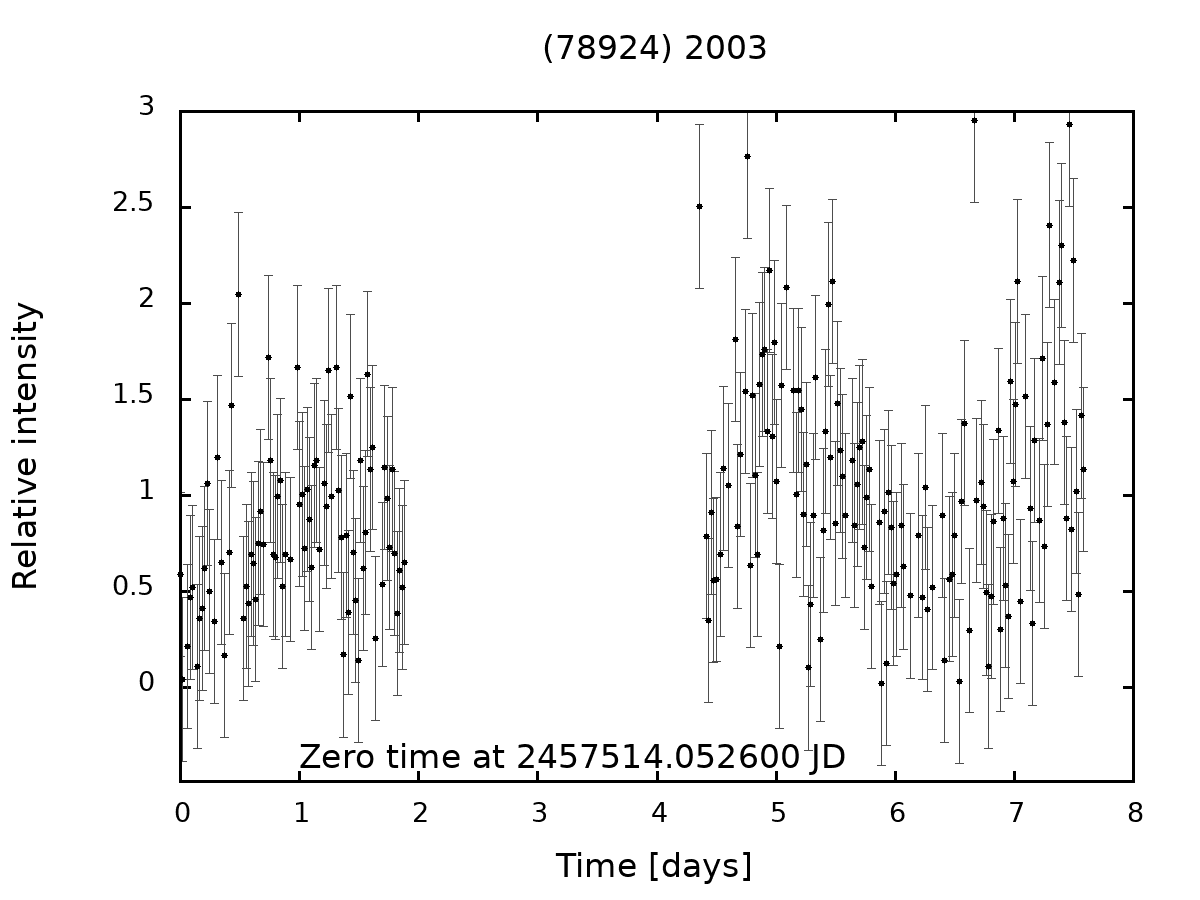}
\end{center}
\caption{\label{fig4a} Light curves of asteroids with lower limit on rotation period.}
\end{figure*}

\begin{figure*}
\begin{center}
\includegraphics[width=0.327\columnwidth]{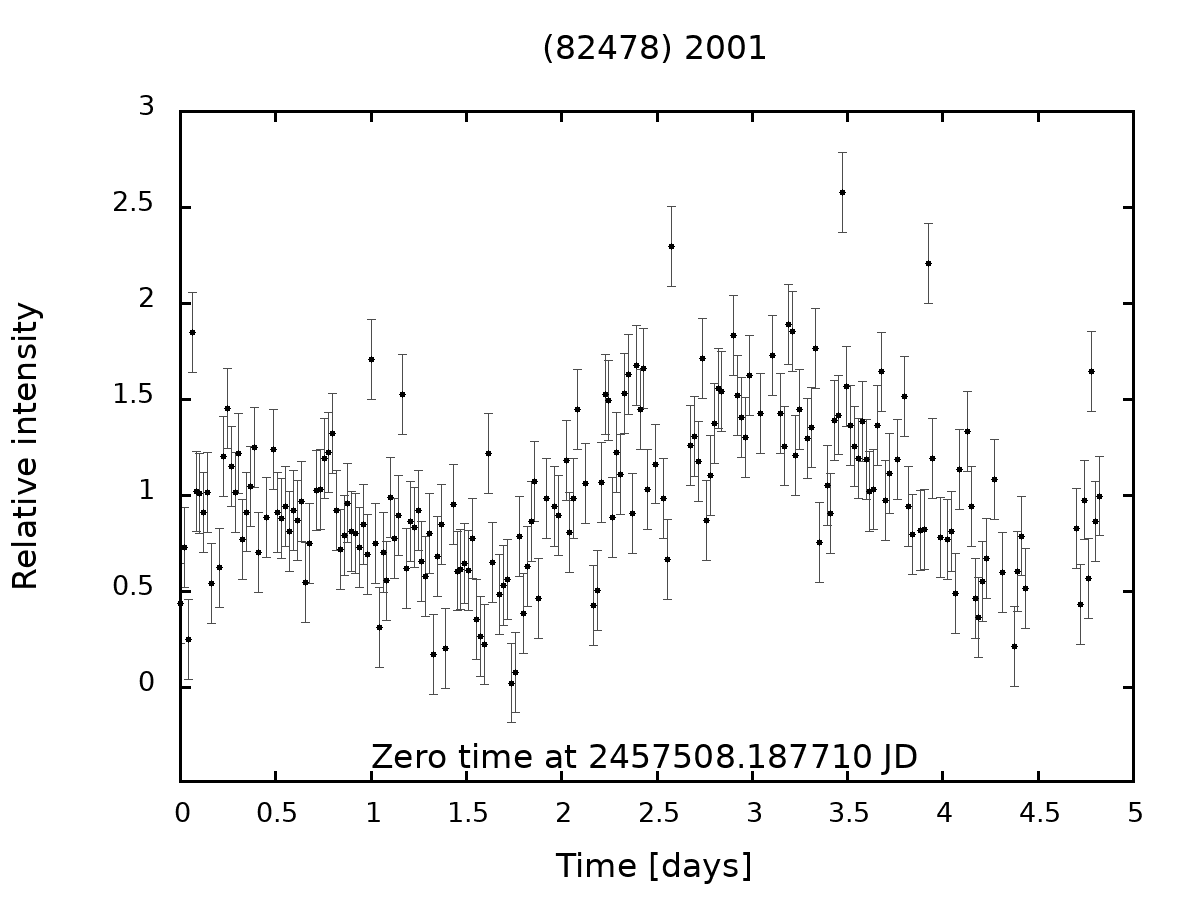}
\includegraphics[width=0.327\columnwidth]{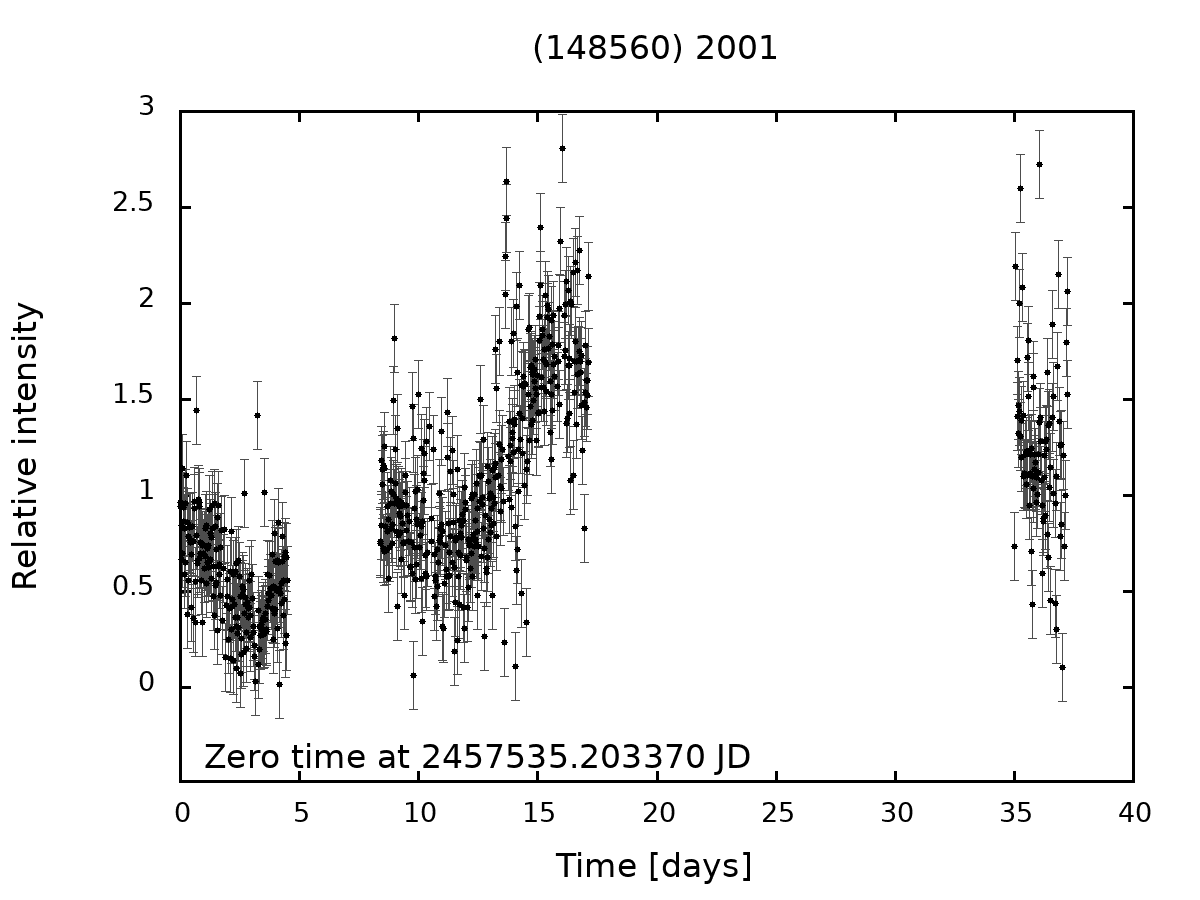}
\includegraphics[width=0.327\columnwidth]{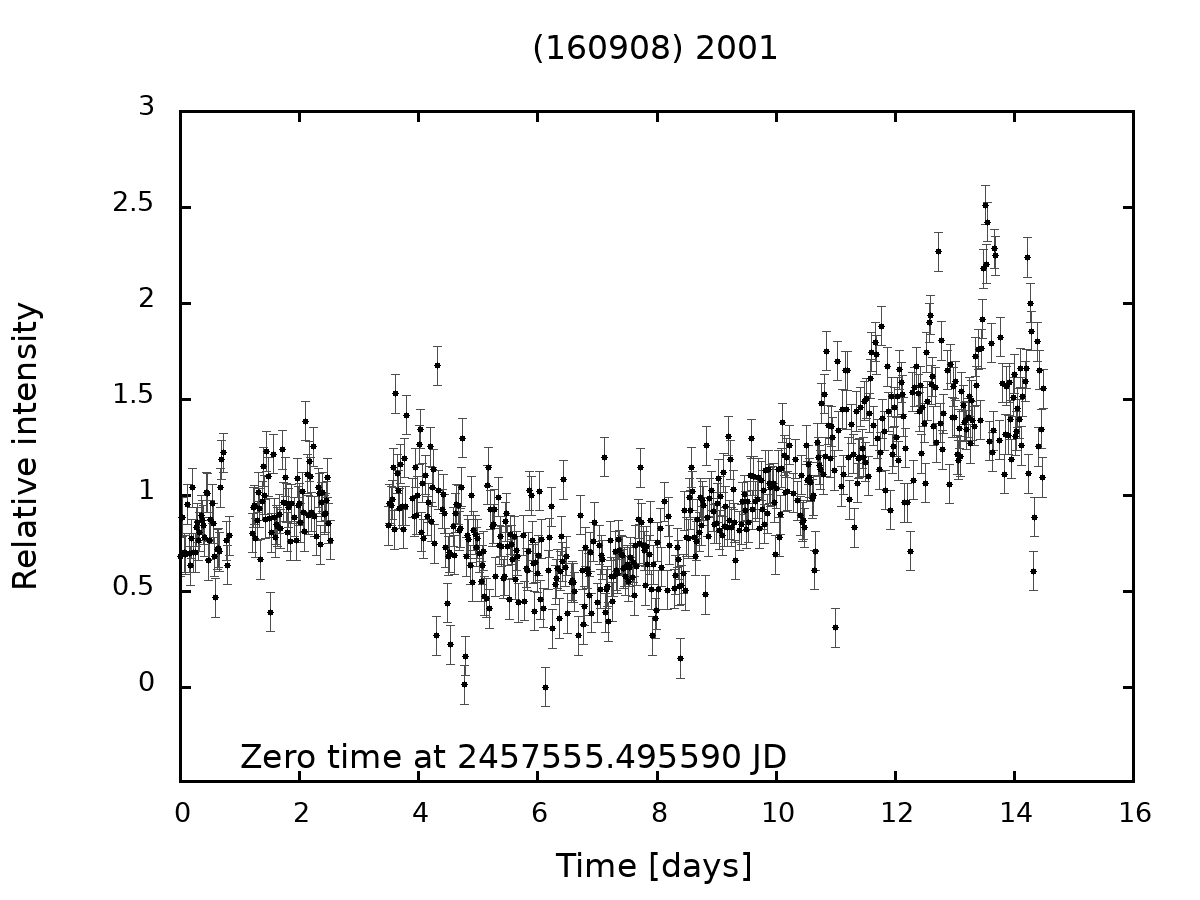}
\includegraphics[width=0.327\columnwidth]{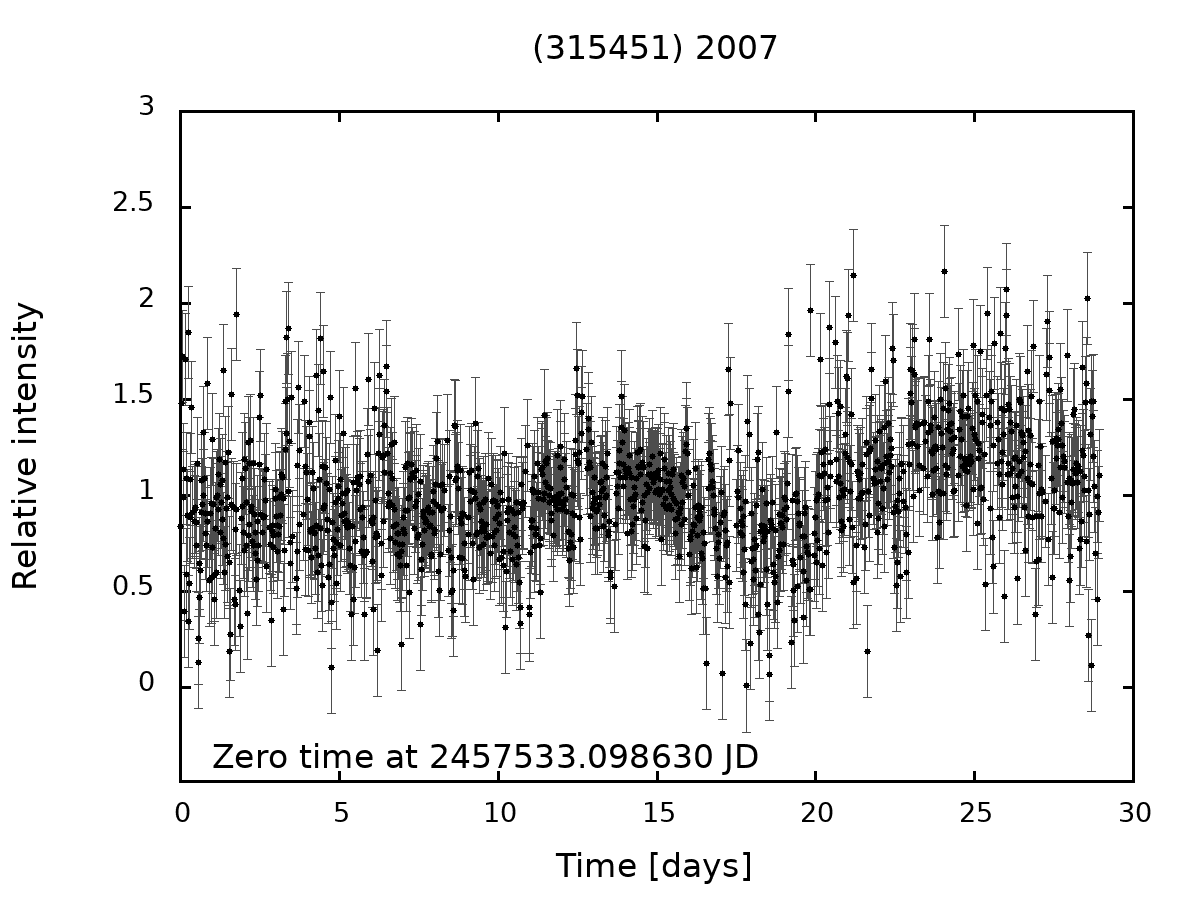}
\includegraphics[width=0.327\columnwidth]{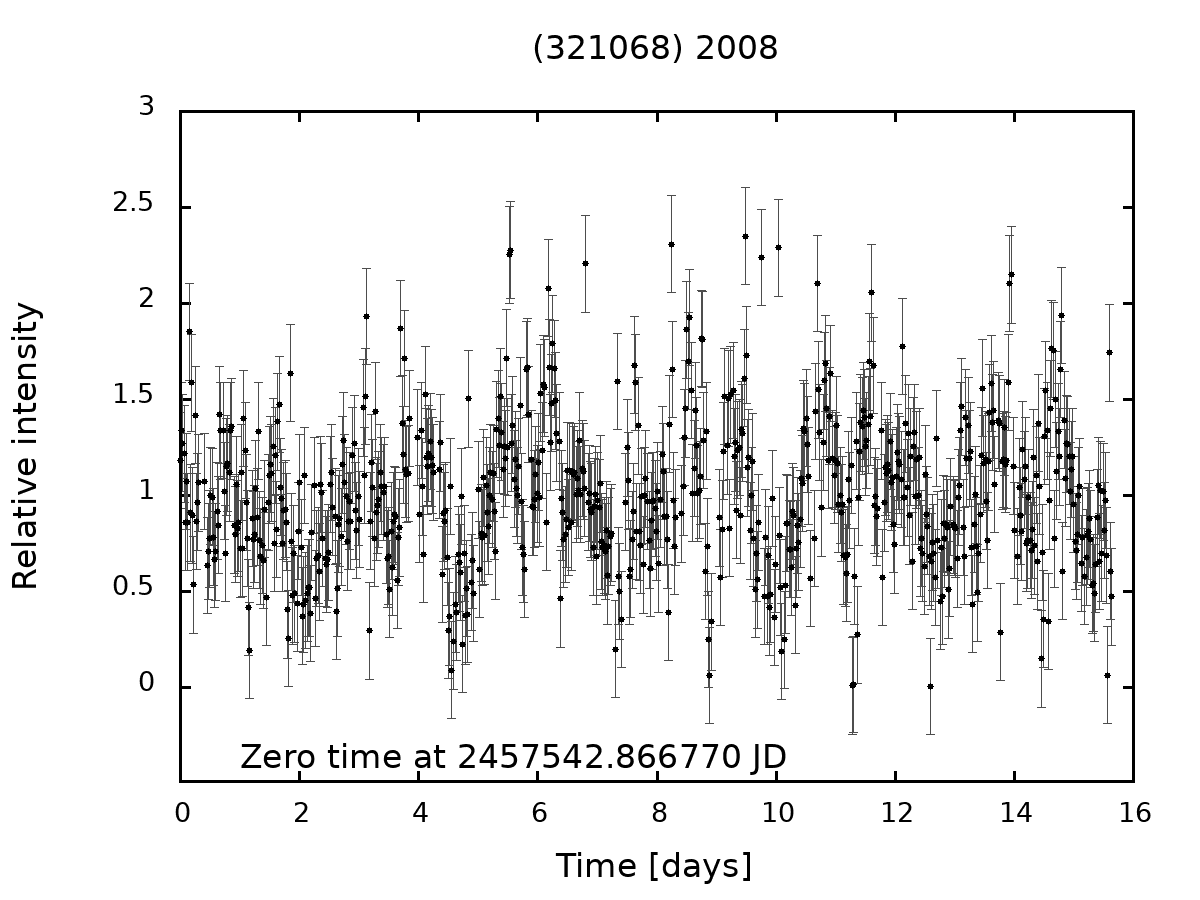}
\includegraphics[width=0.327\columnwidth]{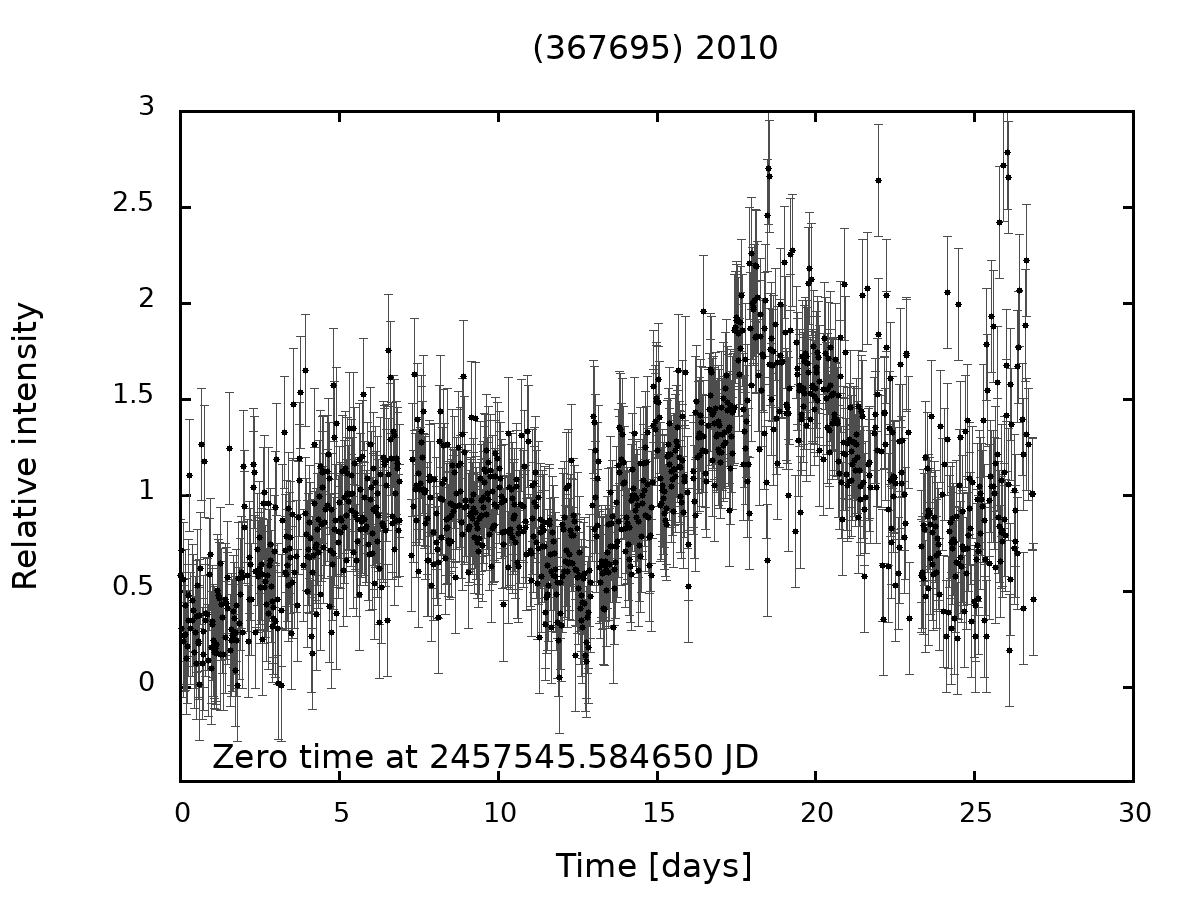}
\includegraphics[width=0.327\columnwidth]{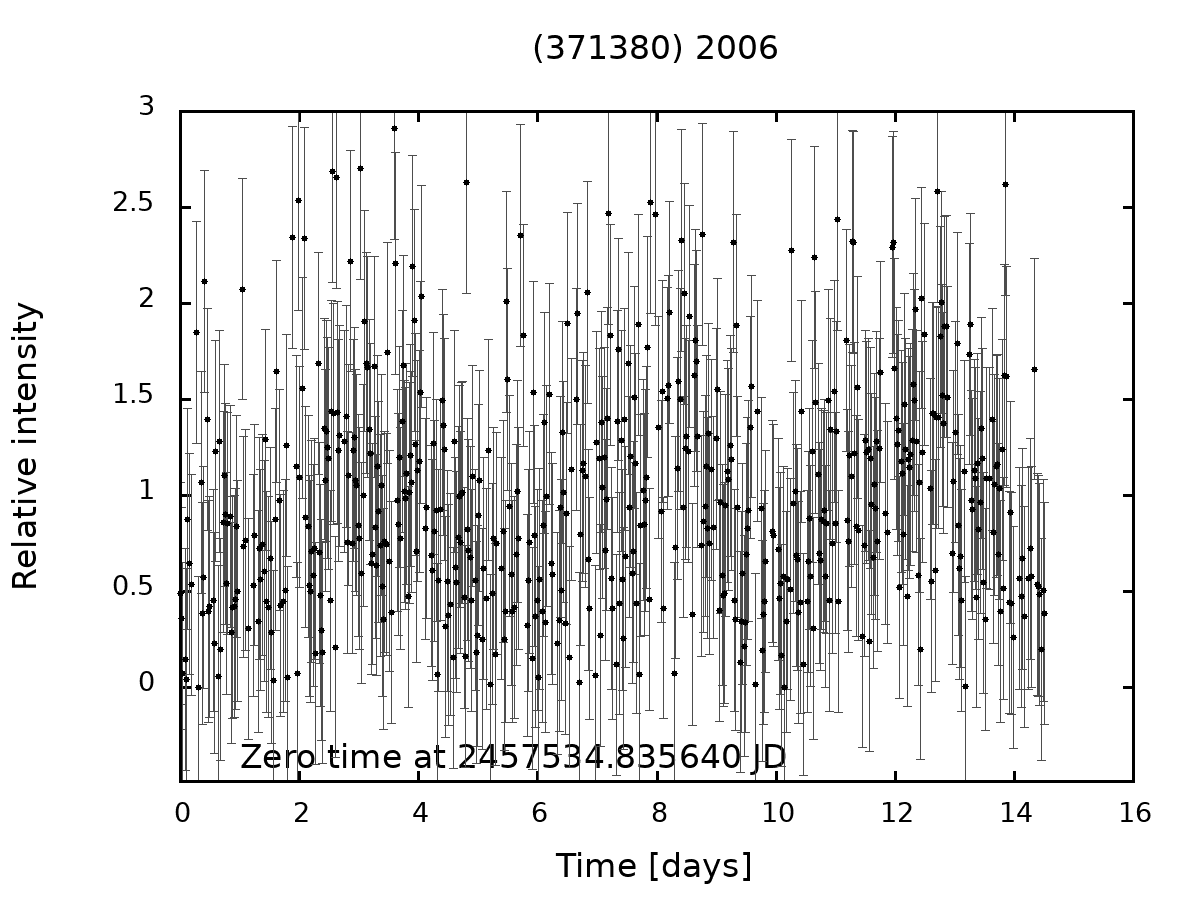}
\includegraphics[width=0.327\columnwidth]{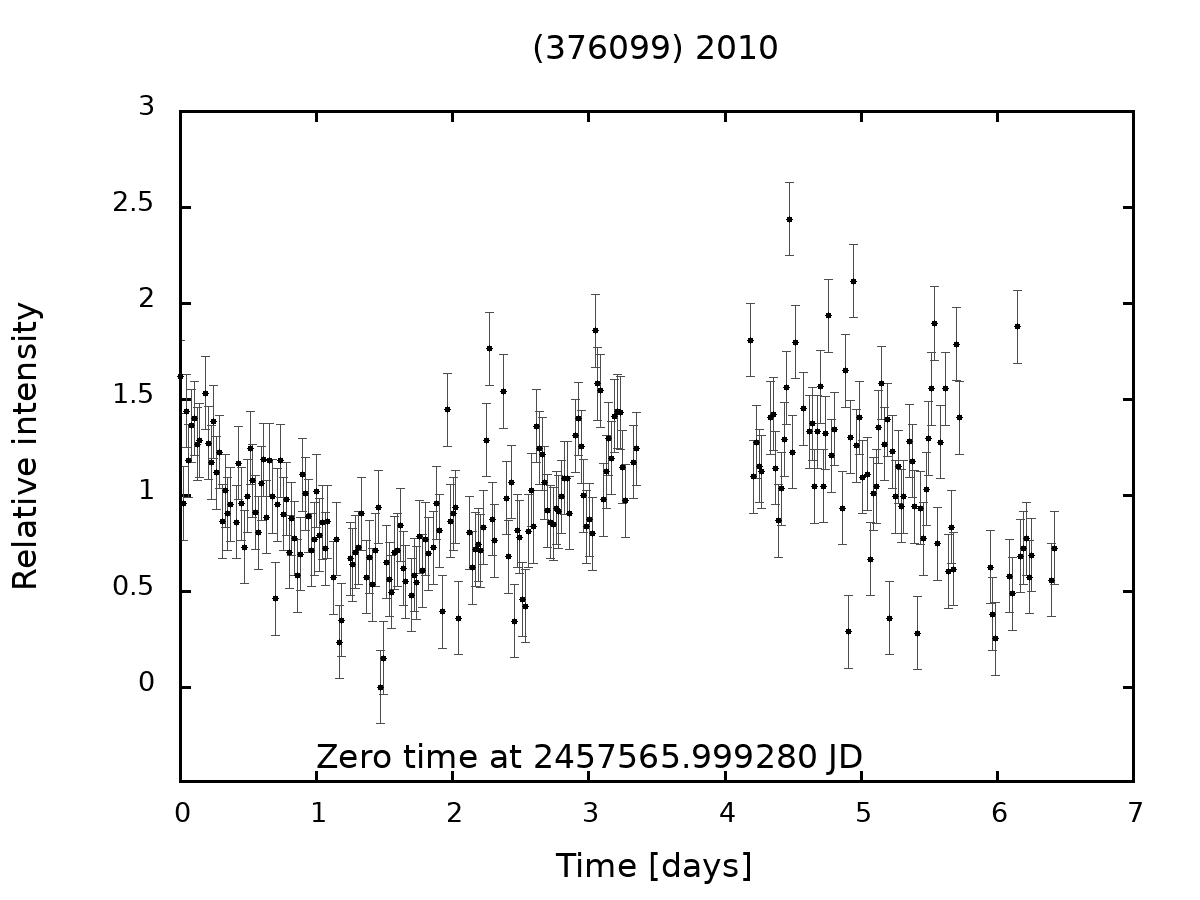}
\end{center}
\caption{\label{fig4b} Light curves of asteroids with lower limit on rotation period -- \textit{continued}.}
\end{figure*}

\begin{figure}
\begin{center}
\includegraphics[width=\columnwidth]{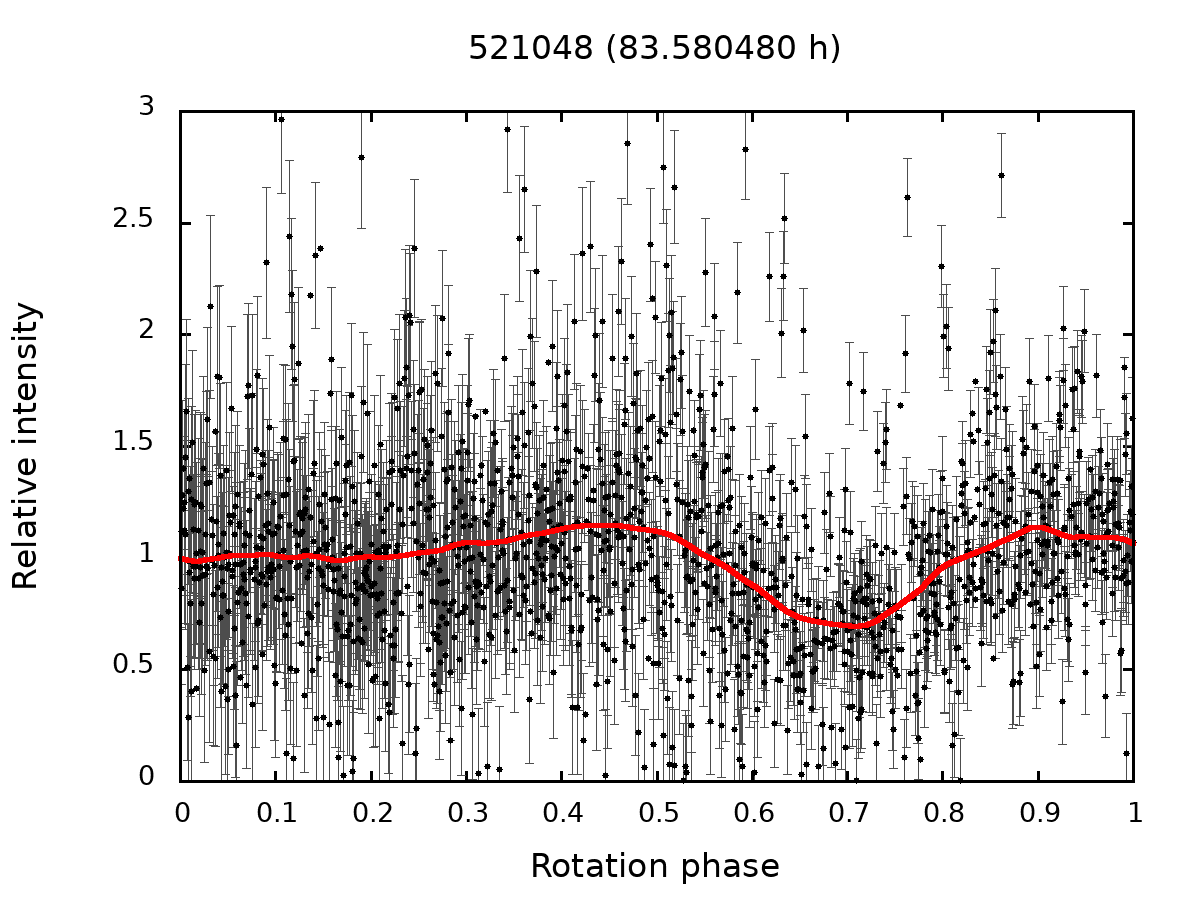}

\end{center}
\caption{\label{occ}{ Lightcurve of asteroid 521048 with minima indicating possible occultation events. The peak occurs every $83.58$ h.  }}
\end{figure}

Moreover, we have selected a group of 20 asteroids for which the data contain a clear trend indicating that we can expect long rotation periods (longer than $100$ h) for these objects, but the time of the observations was too short to cover the whole rotation period or to give certain results. A very good example here is the asteroid 4635, which has a known rotation period around $117$ h. Our results suggest similar value, but further observations would be needed to confirm it.
 In the group of slow rotators there are four asteroids with rotation periods known from LCDB. These are 3233, which has an extremely long $P=888$ h, 4472 with $P=20$ h (but with quality code 1+), which makes our determination more probable,  4635 mentioned above, and 33679 where $P=12$ h is in contrary with our $P>233$ h.
Lightcurves of all these objects are shown in Figures~\ref{fig4a} and \ref{fig4b}. In Table \ref{tab:all1} we provide the data for these asteroids with the minimum estimated period from available dataset.
A very interesting asteroid in the group of slow rotators is 521048, with almost zero amplitude, but displaying a decrease of the brightness every $83$ h. This feature may suggest the occultation by a smaller companion. However, the number of data points for this object is not enough to confirm the feature. Figure~\ref{occ} shows the lightcurve of this object.

\section{Discussion and conclusions}

Using the data obtained by the \textit{Kepler} satellite during \textit{K2}C9 we have extracted the photometry of the asteroids passing in the spacecraft field of view. Our aim was to determine for the first time the rotation periods for a large number of asteroids. From 1395 investigated objects it was possible to obtain photometric data for 1026 asteroids. Among them there were objects for which the number of data points did not allow to estimate the reliable rotation periods, or objects for which amplitude of brightness changes was too small to extract signal from the noise. However, we were able to calculate rotation periods for 188 asteroids, with the $S/N$ ratio high enough to ensure trustworthy results. Our results were confirmed by means of two numerical codes (FNPEAKS and AOV) and results from two types of photometry (aperture-like and PRF-like) were taken into account. 
The uncertainty of period determination was provided for each object. This gives valuable information about the reliability of
our estimations.
As was already mentioned, we have taken into account the objects for which the PRF-like and the aperture-like photometry give the same results within calculated uncertainty.
Moreover, there was a selected group of 20 asteroids for which we expect very long rotation periods ($>100$ h) and one candidate for an eclipsing binary object.

\begin{figure}
\begin{center}
\includegraphics[width=0.49\columnwidth]{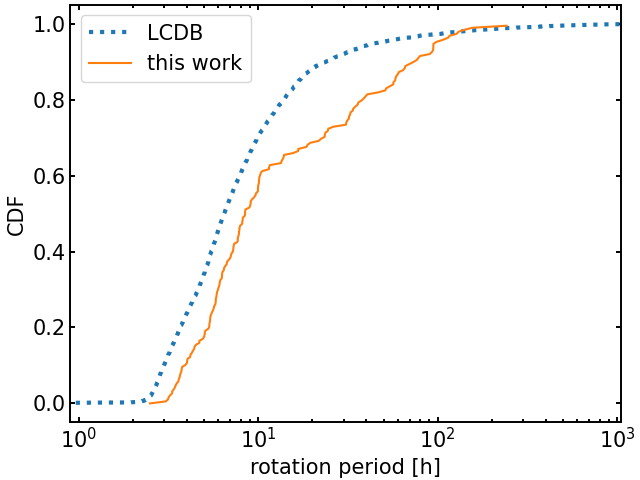}
\includegraphics[width=0.49\columnwidth]{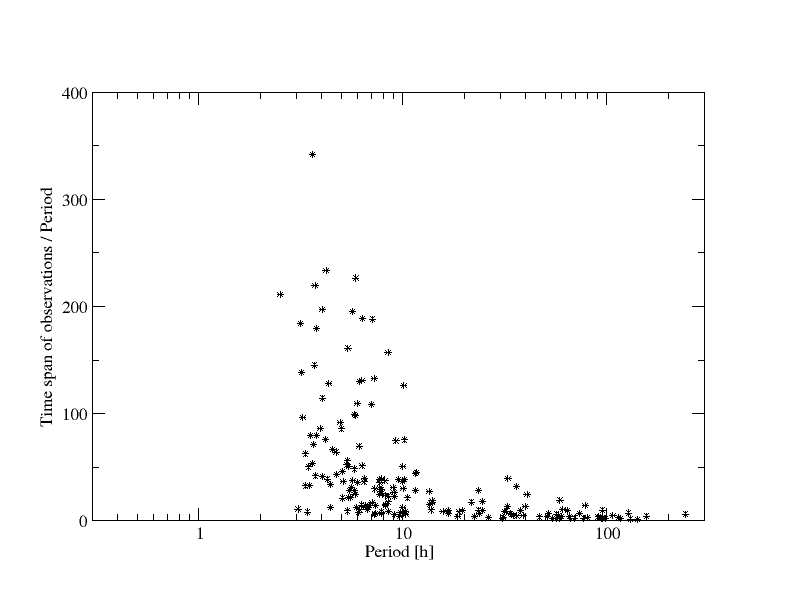}
\end{center}
\caption{\label{nice}{Left: the cumulative distribution function of rotation period. Right: ratio of the time span of the observation to the rotation period ratio vs. rotation period.}}
\end{figure}

We have compared the distribution of rotation periods determined in this paper with the data from LCDB. 
The LCDB sample was limited to objects with a quality code $2+$ or better and with a semi-major axis in the same range as our sample, i.e., $1.14~\mathrm{AU} < a < 5.29~\mathrm{AU}$.
In left panel of Figure~\ref{nice}, we show the cumulative distribution functions of both samples. Our sample clearly lacks the objects with $P<3$ h.
This is partly caused by the length of a single exposure, but also the cutoff at $~2.2$ h from the rotational break-up limit, which reduces observed amplitudes for the shortest period objects. 
The spin barrier is an evidence that the asteroids are mostly loosely bound, gravity-dominated aggregates. The observational and theoretical limits for spin barrier for asteroids close to $2.2$ h was studied in detail in \cite{Pravec, Holsapple}.
Both samples have similar percentage of objects in $3~\mathrm{h} < P < 10~\mathrm{h}$ range. At $P = 10$ h number of objects in our sample drops. Between $\approx30$ h and $\approx150$ h our detection efficiency is larger than the LCDB one. The smaller number of objects in $10~\mathrm{h} < P < 30~\mathrm{h}$ range 
can be due to the fact that faster rotating objects are covered many times during the time span of the observations, which gives better $S/N$ and higher number of measured rotation periods. This is confirmed in right panel of Figure~\ref{nice}, where we show the ratio of the time span of the observations to the rotation period. For very slow rotators the lightcurve is covered only a few times during the time span of the observations. Thus, the most common rotation periods are below $10$ h. 
This result is in agreement with \citep{Pal_tess} who showed that the number of asteroids from \textit{K2} with long rotation period is underestimated. The time span of observations varies significantly between different objects.
Nevertheless, we have selected also the group of asteroids displaying features that indicate long rotation periods, but for which we did not have enough data to determine the exact value. These objects are listed in Table \ref{tab:all1}, where we provide minimal estimated period
for each object. These periods turned out to be very long ($P>100$ h). 

All results presented in this paper are summarized in Tables \ref{tab:results1}-\ref{tab:all1} and the data will be publicly available in CDS  after acceptance of the paper. 
At the end it is worth mentioning that analysing the uncertainty of rotation periods provided for each object we can see that the estimated error is always better than $7\%$ of the rotation period, which indicate that the precision of the calculations is relatively high.

\acknowledgments
Work by RP was supported by Polish National Agency for Academic Exchange grant ``Polish Returns 2019.''
This paper includes data collected by the {\it K2} mission and obtained from the MAST data archive at the Space Telescope Science Institute (STScI). Funding for the {\it K2} mission is provided by the NASA Science Mission Directorate. STScI is operated by the Association of Universities for Research in Astronomy, Inc., under NASA contract NAS 5–26555.

\facilities{K2}

\software{
MCPM \citep{Poleski_method}\footnote{\url{https://github.com/CPM-project/MCPM}},
FNPEAKS \footnote{\url{http://helas.astro.uni.wroc.pl/deliverables.php?lang=pl&active=fnpeaks}},
EPHEMD \citep{Molnar18}, 
AOV\footnote{\url{https://users.camk.edu.pl/alex/\#software}},
JPL ISPY\footnote{\url{https://ssd.jpl.nasa.gov/x/ispy.html}},
JPL Horizons \footnote{\url{https://ssd.jpl.nasa.gov/?horizons}}. 
}


\startlongtable
\begin{deluxetable*}{ccccc}
\tablecaption{List of asteroids with determined rotation periods. The first column shows the number and the name of asteroid, the second number of data points that were for our disposal, the third length of observation, the fourth $S/N$ ratio, and the fifth determined rotation period together with estimated uncertainty. In this Table, we include objects with the most certain rotation period determination.\label{tab:results1}}
\tablewidth{0pt}
\tablehead{
\colhead{Asteroid} & \colhead{Number of} &\colhead{Time span }& \colhead{$S/N$} & \colhead{Period}   \\
\colhead{number} & \colhead{data points}&\colhead{of observations [d]} & \colhead{} & \colhead{[h]}  
}
\decimalcolnumbers
\startdata
(271) Penthesilea & 131 & 2.90 & 5.8 & $18.58\pm{0.55}$ \\ 
(1804) Chebotarev & 882 & 19.19 & 24.0 & $4.0241\pm{0.0025}$ \\ 
(1920) Sarmiento & 301 & 6.91 & 6.5 & $4.043\pm{0.011}$ \\ 
(3414) Champollion & 290 & 11.46 & 5.9 & $5.2811\pm{0.0094}$ \\ 
(3751) Kiang & 127 & 4.84 & 6.6 & $8.212\pm{0.058}$ \\ 
(4631) Yabu & 205 & 4.35 & 12.9 & $7.354\pm{0.054}$ \\ 
(5026) Martes & 214 & 6.25 & 12.5 & $4.425\pm{0.012}$ \\ 
(5114) Yezo & 964 & 23.07 & 15.2 & $4.3326\pm{0.0025}$ \\ 
(5979) 1992 XF & 897 & 51.27 & 27.9 & $3.60185\pm{0.00059}$ \\ 
(6825) Irvine & 291 & 8.07 & 14.1 & $3.6148\pm{0.0064}$ \\ 
(7310) 1995 OL1 & 100 & 2.10 & 8.9 & $7.31\pm{0.14}$ \\ 
(8447) Cornejo & 160 & 3.82 & 6.4 & $6.279\pm{0.057}$ \\ 
(9213) 1995 UX5 & 285 & 6.83 & 11.6 & $36.19\pm{0.79}$ \\ 
(9582) 1990  EL7& 276 & 6.74 & 10.5 & $5.802\pm{0.020}$ \\ 
(9780) Bandersnatch & 185 & 5.01 & 5.8 & $7.052\pm{0.040}$ \\ 
(10125) Stenkyrka & 253 & 7.27 & 10.4 & $5.600\pm{0.020}$ \\ 
(10446) Siegbahn & 234 & 8.89 & 3.9 & $5.680\pm{0.017}$ \\ 
(10688) Haghighipour & 212 & 9.85 & 7.5 & $9.042\pm{0.044}$ \\ 
(10727) Akitsushima & 150 & 3.41 & 7.9 & $6.690\pm{0.065}$ \\ 
(10817) 1993 FR44 & 1280 & 49.74 & 30.2 & $6.3230\pm{0.0019}$ \\ 
(11982) 1995 UF6 & 137 & 5.72 & 6.1 & $13.74\pm{0.19}$ \\ 
(13229) Echion & 1355 & 55.32 & 13.9 & $8.4509\pm{0.0038}$ \\ 
(14658) 1999 AC10 & 921 & 46.14 & 27.2 & $5.6609\pm{0.0019}$ \\ 
(15517) 1999 VS113 & 105 & 3.04 & 4.6 & $8.497\pm{0.099}$ \\ 
(17966) 1999 JS43 & 616 & 47.10 & 23.2 & $78.10\pm{0.49}$ \\ 
(18071) 2000 BA27 & 1369 & 48.02 & 18.8 & $35.920\pm{0.072}$ \\ 
(19742) 2000 AS162 & 1433 & 33.90 & 8.4 & $3.7063\pm{0.0010}$ \\ 
(20694) 1999 VT82 & 278 & 8.64 & 5.1 & $3.3250\pm{0.0048}$ \\ 
(20802) 2000 SR179 & 294 & 9.50 & 4.6 & $7.824\pm{0.026}$ \\ 
(21195) 1994 PK4 & 847 & 23.79 & 10.9 & $5.7751\pm{0.0040}$ \\ 
(22810) Rawat & 234 & 8.42 & 5.6 & $4.711\pm{0.010}$ \\ 
(24002) 1999 RR35 & 864 & 33.19 & 19.2 & $6.1147\pm{0.0043}$ \\ 
(24272) 1999 XE165 & 340 & 8.03 & 14.0 & $33.92\pm{0.57}$ \\ 
(25353) 1999 RB210 & 177 & 7.36 & 5.6 & $34.89\pm{0.71}$ \\ 
(29415) 1996 XU5 & 298 & 6.52 & 10.0 & $16.75\pm{0.18}$ \\ 
(30465) 2000 OY13 & 292 & 9.36 & 5.7 & $10.532\pm{0.042}$ \\ 
(30617) 3068 P-L & 1293 & 34.41 & 20.4 & $6.2960\pm{0.0032}$ \\ 
(30892) 1993 FR18 & 315 & 8.11 & 6.3 & $8.481\pm{0.034}$ \\ 
(34254) Mihirpatel & 549 & 12.49 & 10.9 & $10.021\pm{0.027}$ \\ 
(35751) 1999 GE36 & 611 & 13.43 & 12.0 & $90.5\pm{2.0}$ \\ 
(36324) 2000 LT27 & 299 & 7.79 & 12.3 & $5.086\pm{0.015}$ \\ 
(36662) 2000 QT208 & 715 & 18.82 & 12.8 & $4.9186\pm{0.0042}$ \\ 
(36700) 2000 RT17 & 424 & 11.59 & 10.5 & $94.8\pm{2.9}$ \\ 
(37104) 2000 UP99 & 170 & 5.03 & 5.0 & $5.529\pm{0.025}$ \\ 
(38680) 2000 QM2 & 199 & 6.62 & 7.8 & $66.1\pm{2.6}$ \\ 
(41613) 2000 SP144 & 502 & 11.67 & 6.8 & $3.5284\pm{0.0035}$ \\ 
(41739) 2000 UP112 & 427 & 12.57 & 8.8 & $4.7032\pm{0.0069}$ \\ 
(43629) 2002 CG234 & 246 & 6.17 & 6.4 & $8.456\pm{0.051}$ \\ 
(44957) 1999 VG78 & 253 & 5.89 & 6.3 & $5.820\pm{0.023}$ \\ 
(48822) 1997 WY35 & 317 & 12.40 & 5.5 & $3.7543\pm{0.0044}$ \\ 
(49550) 1999 CO84 & 315 & 9.65 & 8.2 & $5.035\pm{0.010}$ \\ 
(49659) 1999 JC118 & 220 & 5.11 & 7.5 & $8.227\pm{0.061}$ \\ 
(50971) 2000 GP88 & 136 & 6.52 & 8.3 & $3.7394\pm{0.0078}$ \\ 
(51138) 2000 HC45 & 879 & 20.99 & 21.0 & $11.499\pm{0.020}$ \\ 
(51260) 2000 JZ62 & 827 & 18.31 & 9.6 & $3.1752\pm{0.0018}$ \\ 
(52621) 1997 VW4 & 330 & 7.19 & 7.7 & $3.4403\pm{0.0062}$ \\ 
(53402) 1999 JG119 & 143 & 4.09 & 9.0 & $6.841\pm{0.050}$ \\ 
(56341) 1999 XS221 & 351 & 14.63 & 10.6 & $56.78\pm{0.85}$ \\ 
(61346) 2000 PD8 & 236 & 8.21 & 6.1 & $51.2\pm{1.3}$ \\ 
(62176) 2000 SJ36 & 247 & 7.93 & 5.1 & $7.710\pm{0.031}$ \\ 
(64501) 2001 VZ65 & 547 & 17.70 & 6.8 & $6.0965\pm{0.0066}$ \\ 
(64948) 2001 YH124 & 277 & 8.42 & 10.6 & $13.515\pm{0.076}$ \\ 
(65354) 2002 NG43 & 299 & 7.07 & 5.7 & $94.5\pm{5.1}$ \\ 
(66833) 1999 UP45 & 169 & 6.91 & 8.5 & $4.272\pm{0.013}$ \\ 
(69567) 1998 BC8 & 332 & 9.69 & 6.0 & $23.54\pm{0.18}$ \\ 
(70361) 1999 RK189 & 326 & 10.16 & 7.5 & $116.5\pm{5.3}$ \\ 
(73062) 2002 FE10 & 114 & 2.57 & 5.5 & $6.155\pm{0.076}$ \\ 
(73643) 1978 UA5 & 916 & 21.07 & 8.8 & $9.991\pm{0.014}$ \\ 
(73653) 1981 EN6 & 224 & 4.88 & 12.2 & $56.7\pm{2.9}$ \\ 
(74534) 1999 JA & 237 & 5.21 & 11.2 & $9.986\pm{0.083}$ \\ 
(77590) 2001 KM17 & 129 & 3.09 & 6.8 & $5.904\pm{0.055}$ \\ 
(77871) 2001 SC9 & 405 & 8.91 & 17.0 & $94.7\pm{4.2}$ \\ 
(77974) 2002 JD13 & 122 & 3.80 & 6.1 & $31.0\pm{1.3}$ \\ 
(80910) 2000 DL60 & 472 & 10.56 & 14.0 & $6.460\pm{0.014}$ \\ 
(82694) 2001 PB35 & 716 & 24.07 & 22.2 & $5.8660\pm{0.0050}$ \\ 
(83441) 2001 SL56 & 262 & 6.72 & 4.3 & $23.72\pm{0.34}$ \\ 
(84229) 2002 SH15 & 461 & 17.78 & 14.0 & $32.57\pm{0.21}$ \\ 
(84282) 2002 TV23 & 642 & 27.12 & 7.7 & $5.9605\pm{0.0034}$ \\ 
(87630) 2000 RE55 & 1395 & 31.94 & 33.1 & $10.1546\pm{0.0088}$ \\ 
(87633) 2000 RM64 & 322 & 12.49 & 11.5 & $4.4991\pm{0.0051}$ \\ 
(88472) 2001 QW111 & 200 & 6.42 & 8.9 & $5.472\pm{0.024}$ \\ 
(89237) 2001 UD152 & 253 & 9.95 & 7.5 & $79.8\pm{2.7}$ \\ 
(90789) 1994 PP22 & 767 & 17.88 & 16.7 & $5.0069\pm{0.0043}$ \\ 
(93735) 2000 VO48 & 600 & 16.31 & 13.5 & $10.147\pm{0.020}$ \\ 
(95334) 2002 CD118 & 348 & 13.45 & 9.6 & $11.517\pm{0.037}$ \\ 
(95666) 2002 GZ141 & 825 & 32.98 & 11.3 & $4.0154\pm{0.0016}$ \\ 
(97093) 1999 VQ59 & 308 & 7.13 & 6.6 & $69.5\pm{2.9}$ \\ 
(97674) 2000 FJ50 & 347 & 8.03 & 7.9 & $8.169\pm{0.033}$ \\ 
(97811) 2000 OH33 & 318 & 9.11 & 7.5 & $33.70\pm{0.42}$ \\ 
(100034) 1991 PN1 & 329 & 9.73 & 4.3 & $6.447\pm{0.016}$ \\ 
(101902) Gisellaluccone & 302 & 9.22 & 5.8 & $60.1\pm{1.4}$ \\ 
(104944) 2000 JN41 & 203 & 4.45 & 7.7 & $5.050\pm{0.025}$ \\ 
(105357) 2000 QF109 & 542 & 12.98 & 7.0 & $3.2297\pm{0.0026}$ \\ 
(107076) 2001 AM21 & 497 & 15.39 & 4.9 & $13.427\pm{0.042}$ \\ 
(110041) 2001 SJ82 & 281 & 10.83 & 6.6 & $3.6542\pm{0.0045}$ \\ 
(111351) 2001 XZ107 & 700 & 15.20 & 22.7 & $9.494\pm{0.019}$ \\ 
(113994) 2002 UN32 & 435 & 10.07 & 6.4 & $93.1\pm{3.4}$ \\ 
(126139) 2001 YD128 & 771 & 24.13 & 6.1 & $3.1497\pm{0.0013}$ \\ 
(126750) 2002 DX1 & 456 & 11.32 & 8.1 & $7.712\pm{0.019}$ \\ 
(126963) 2002 FG16 & 378 & 8.99 & 5.0 & $7.282\pm{0.023}$ \\ 
(127777) 2003 FT53 & 202 & 4.90 & 5.1 & $5.425\pm{0.031}$ \\ 
(128258) 2003 SF297 & 376 & 12.42 & 4.9 & $5.3482\pm{0.0078}$ \\ 
(132598) 2002 JG142 & 118 & 2.60 & 4.4 & $10.34\pm{0.20}$ \\ 
(140764) 2001 UB122 & 124 & 2.96 & 6.0 & $6.711\pm{0.078}$ \\ 
(141426) 2002 CX6 & 153 & 3.64 & 5.0 & $22.44\pm{0.69}$ \\ 
(141573) 2002 GC132 & 909 & 21.33 & 10.2 & $106.9\pm{1.9}$ \\ 
(141577) 2002 GS149 & 1284 & 47.04 & 17.7 & $58.84\pm{0.18}$ \\ 
(142239) 2002 RJ93 & 131 & 3.56 & 6.5 & $10.27\pm{0.15}$ \\ 
(144875) 2004 OX & 208 & 4.86 & 10.1 & $5.409\pm{0.028}$ \\ 
(152309) 2005 TS95 & 379 & 12.42 & 5.0 & $58.1\pm{1.0}$ \\ 
(164134) 2003 YY48 & 360 & 7.87 & 13.1 & $19.49\pm{0.17}$ \\ 
(166816) 2002 VO88 & 206 & 8.54 & 5.9 & $7.950\pm{0.034}$ \\ 
(169114) 2001 OK47 & 411 & 12.85 & 7.1 & $7.880\pm{0.015}$ \\ 
(169311) 2001 TV81 & 1212 & 27.34 & 28.4 & $23.490\pm{0.058}$ \\ 
(170144) 2003 AN81 & 395 & 9.81 & 5.5 & $13.947\pm{0.078}$ \\ 
(170972) 2005 CS22 & 400 & 9.36 & 6.8 & $7.774\pm{0.023}$ \\ 
(172778) 2004 EE64 & 419 & 13.20 & 9.2 & $4.1813\pm{0.0043}$ \\ 
(173509) 2000 UK50 & 432 & 11.32 & 12.4 & $5.4010\pm{0.0088}$ \\ 
(174006) 2001 YT7 & 1779 & 53.64 & 10.8 & $32.709\pm{0.049}$ \\ 
(196663) 2003 SL44 & 1581 & 38.28 & 22.7 & $127.8\pm{1.1}$ \\ 
(199543) 2006 DL210 & 331 & 11.44 & 4.8 & $31.35\pm{0.31}$ \\ 
(211183) 2002 JF113 & 113 & 3.11 & 9.3 & $26.3\pm{1.1}$ \\ 
(212937) 2008 UD291 & 411 & 9.87 & 6.8 & $65.7\pm{1.6}$ \\ 
(228119) 2008 TM136 & 1749 & 61.98 & 15.4 & $242.4\pm{2.7}$ \\ 
(228780) 2002 XS84 & 282 & 19.97 & 15.1 & $73.17\pm{0.87}$ \\ 
(240067) 2001 XH225 & 297 & 7.70 & 7.4 & $38.83\pm{0.91}$ \\ 
(258914) 2002 RP5 & 1143 & 25.03 & 25.8 & $64.22\pm{0.49}$ \\ 
(262257) 2006 SF326 & 661 & 18.47 & 4.8 & $24.62\pm{0.10}$ \\ 
(269046) 2007 FQ38 & 408 & 11.65 & 7.2 & $8.947\pm{0.030}$ \\ 
(271713) 2004 RU200 & 682 & 40.85 & 10.1 & $4.1974\pm{0.0028}$ \\ 
(280384) 2003 UW118 & 959 & 31.72 & 8.8 & $7.0038\pm{0.0055}$ \\ 
(288366) 2004 CB15 & 608 & 14.86 & 7.4 & $37.72\pm{0.33}$ \\ 
(321066) 2008 SF56 & 289 & 8.93 & 4.7 & $5.999\pm{0.014}$ \\ 
(323343) 2003 US267 & 528 & 12.30 & 7.9 & $7.676\pm{0.016}$ \\ 
(325465) 2009 QM52 & 473 & 13.45 & 9.6 & $6.304\pm{0.011}$ \\ 
(335611) 2006 EW71 & 648 & 28.20 & 7.1 & $3.7687\pm{0.0014}$ \\ 
(342867) 2008 YP34 & 1636 & 55.50 & 11.3 & $5.8784\pm{0.0018}$ \\ 
(344209) 2001 QZ265 & 838 & 55.60 & 7.4 & $7.0918\pm{0.0039}$ \\ 
(351327) 2004 VR75 & 278 & 6.74 & 7.4 & $46.6\pm{1.3}$ \\ 
(371411) 2006 SF34 & 395 & 11.69 & 9.9 & $5.807\pm{0.011}$ \\ 
(371691) 2007 DD59 & 843 & 36.11 & 10.2 & $5.3766\pm{0.0020}$ \\ 
(376966) 2002 JW57 & 363 & 8.34 & 7.6 & $130.8\pm{8.6}$ \\ 
(379752) 2011 GN82 & 507 & 12.63 & 5.7 & $8.162\pm{0.019}$ \\ 
(382243) 2012 SV39 & 1155 & 39.89 & 9.6 & $7.2166\pm{0.0033}$ \\ 
(418883) 2008 YH88 & 1651 & 39.05 & 15.6 & $94.73\pm{0.70}$ \\ 
(419549) 2010 OZ83 & 638 & 15.76 & 7.8 & $21.75\pm{0.10}$ \\ 
(434707) 2006 CH49 & 975 & 53.23 & 6.7 & $10.1059\pm{0.0051}$ \\ 
(468687) 2009 SL103 & 277 & 7.21 & 5.8 & $98.6\pm{5.8}$ \\ 
(500095) 2012 BT20  & 1277 & 41.30 & 12.9 & $40.56\pm{0.14}$ \\ 
(507894) 2014 SF261 & 302 & 11.97 & 4.9 & $32.08\pm{0.32}$ \\ 
(531563) 2012 TP284 & 356 & 8.46 & 6.3 & $9.094\pm{0.037}$ \\ 
2012 SQ50 & 1120 & 28.71 & 8.8 & $9.2433\pm{0.0082}$ \\ 
2013 VA22 & 672 & 15.29 & 9.3 & $9.979\pm{0.021}$ \\ 
2016 GU189 & 234 & 5.46 & 4.6 & $15.72\pm{0.20}$ \\ 

\enddata
\end{deluxetable*}
\startlongtable
\begin{deluxetable*}{ccccc}
\tablecaption{The same as in the Table \ref{tab:results1} but for asteroids with less certain rotation period determination.\label{tab:results2}}
\tablewidth{0pt}
\tablehead{
\colhead{Asteroid} & \colhead{Number of} &\colhead{Time span }& \colhead{$S/N$} & \colhead{Period}   \\
\colhead{number} & \colhead{data points}&\colhead{of observations [d]} & \colhead{} & \colhead{[h]}  
}
\decimalcolnumbers
\startdata
(2088) Sahlia & 236 & 5.70 & 5.5 & $58.3\pm{2.6}$ \\ 
(2662) Kandinsky & 1001 & 21.91 & 5.7 & $39.68\pm{0.27}$ \\ 
(3989) Odin & 94 & 2.02 & 4.3 & $5.348\pm{0.069}$ \\ 
(4993) Cossard & 85 & 1.78 & 6.4 & $7.27\pm{0.15}$ \\ 
(8655) 1990 QJ1 & 109 & 2.27 & 5.3 & $4.402\pm{0.041}$ \\ 
(18418) Ujibe & 115 & 4.78 & 3.9 & $3.473\pm{0.010}$ \\ 
(19387) 1998 DA2 & 141 & 4.90 & 4.9 & $16.74\pm{0.27}$ \\ 
(19616) 1999 OS3 & 55 & 1.14 & 4.8 & $3.393\pm{0.071}$ \\ 
(23031) 1999 XX7 & 53 & 1.39 & 4.5 & $3.059\pm{0.052}$ \\ 
(28991) 2001 MU27 & 90 & 3.23 & 4.7 & $9.73\pm{0.19}$ \\ 
(29849) 1999 FJ25 & 460 & 10.69 & 4.7 & $13.880\pm{0.061}$ \\ 
(35815) 1999 JO48 & 46 & 1.59 & 4.3 & $9.67\pm{0.52}$ \\ 
(37393) 2001 XF24 & 82 & 2.88 & 6.3 & $10.10\pm{0.18}$ \\ 
(40550) 1999 RM113 & 83 & 3.74 & 4.8 & $6.565\pm{0.060}$ \\ 
(42886) 1999 RL150 & 112 & 4.01 & 5.1 & $54.1\pm{3.0}$ \\ 
(53234) 1999 CU117 & 961 & 22.01 & 5.8 & $2.49905\pm{0.00079}$ \\ 
(67793) 2000 UE102 & 966 & 22.23 & 7.7 & $3.6740\pm{0.0018}$ \\ 
(71960) 2000 WV114 & 950 & 21.80 & 4.8 & $11.640\pm{0.018}$ \\ 
(89347) 2001 VS66 & 91 & 2.29 & 4.3 & $9.09\pm{0.21}$ \\ 
(105616) 2000 RF102 & 96 & 2.19 & 4.0 & $7.83\pm{0.16}$ \\ 
(109118) 2001 QH42 & 112 & 2.66 & 3.9 & $7.87\pm{0.14}$ \\ 
(116083) 2003 WA125 & 185 & 4.50 & 4.5 & $3.320\pm{0.011}$ \\ 
(122537) 2000 QX218 & 118 & 2.53 & 4.1 & $30.9\pm{1.9}$ \\ 
(169330) 2001 TZ162 & 623 & 14.12 & 5.7 & $3.9268\pm{0.0035}$ \\ 
(172029) 2001 VP39 & 239 & 9.89 & 3.6 & $24.48\pm{0.21}$ \\ 
(179809) 2002 TZ72 & 616 & 13.49 & 5.2 & $51.7\pm{1.4}$ \\ 
(195409) 2002 GK38 & 89 & 1.94 & 4.0 & $6.06\pm{0.12}$ \\ 
(222662) 2001 XR221 & 295 & 6.40 & 4.4 & $18.86\pm{0.23}$ \\ 
(253698) 2003 UP271 & 654 & 27.73 & 4.6 & $156.3\pm{2.2}$ \\ 
(327822) 2006 WU18 & 638 & 14.63 & 7.9 & $114.3\pm{3.3}$ \\ 
(328945) 2010 VR62 & 547 & 26.75 & 4.4 & $60.77\pm{0.51}$ \\ 
(346145) 2007 VQ237 & 251 & 6.23 & 4.4 & $141.4\pm{11.2}$ \\ 
2013 TX103 & 348 & 7.77 & 4.4 & $76.5\pm{3.3}$ \\ 
\enddata
\end{deluxetable*}

\startlongtable
\begin{deluxetable*}{cccc}
\tablecaption{List of asteroids for which we expect long rotation periods ($P>100~\mathrm{h}$), but the observing time span was too short to determine the exact value. \label{tab:all1}}
\tablewidth{0pt}
\tablehead{
\colhead{Asteroid} & \colhead{Number of} &\colhead{Time span } &\colhead{Estimated minimal }\\
\colhead{number} & \colhead{data points}&\colhead{of observations [d]} &\colhead{period [h]}}
\decimalcolnumbers
\startdata
(3233) Krisbarons & 549 & 11.65 & 800\\ 
(4472) Navashin & 174 & 4.76 & 100\\ 
(4635) Rimbaud & 156 & 6.03 & 125\\ 
(15001) Fuzhou & 175 & 3.80 & 170\\ 
(24504) 2001 AD45 & 264 & 6.44 & 300\\ 
(24961) 1997 TO24 & 505 & 11.16 & 200\\ 
(33679) 1999 JY107 & 307 & 10.48 & 233\\ 
(35085) 1990 SL11 & 199 & 6.05 & 128\\ 
(37326) 2001 QA79 & 257 & 8.48 & 164\\ 
(64678) 2001 XQ68 & 1564 & 40.38 & 1200\\ 
(66197) 1999 BO6 & 369 & 8.38 & 700\\ 
(78924) 2003 SD111 & 217 & 7.58 & 100\\ 
(82478) 2001 OS25 & 193 & 4.82 & 115\\ 
(148560) 2001 QP204 & 651 & 37.23 & 480\\ 
(160908) 2001 UM138 & 586 & 14.49 & 450\\ 
(315451) 2007 XF9 & 1250 & 28.92 & 500\\ 
(321068) 2008 SZ56 & 695 & 15.63 & 129\\ 
(367695) 2010 RX113 & 1136 & 26.85 & 550\\ 
(371380) 2006 QU122 & 630 & 14.53 & 220\\ 
(376099) 2010 VP184 & 229 & 6.44 & 220\\ 

\enddata
\end{deluxetable*}

\bibliography{literature}{}

\begin{thebibliography}{}
\expandafter\ifx\csname natexlab\endcsname\relax\def\natexlab#1{#1}\fi
\providecommand{\url}[1]{\href{#1}{#1}}
\providecommand{\dodoi}[1]{doi:~\href{http://doi.org/#1}{\nolinkurl{#1}}}
\providecommand{\doeprint}[1]{\href{http://ascl.net/#1}{\nolinkurl{http://ascl.net/#1}}}
\providecommand{\doarXiv}[1]{\href{https://arxiv.org/abs/#1}{\nolinkurl{https://arxiv.org/abs/#1}}}

\bibitem[{{Bartczak} \& {Dudzi{\'n}ski}(2019)}]{Bartczak}
{Bartczak}, P., \& {Dudzi{\'n}ski}, G. 2019, \mnras, 485, 2431,
  \dodoi{10.1093/mnras/stz300}

\bibitem[{{Borucki} {et~al.}(2010){Borucki}, {Koch}, {Basri}, {Batalha},
  {Brown}, {Caldwell}, {Caldwell}, {Christensen-Dalsgaard}, {Cochran},
  {DeVore}, {Dunham}, {Dupree}, {Gautier}, {Geary}, {Gilliland}, {Gould},
  {Howell}, {Jenkins}, {Kondo}, {Latham}, {Marcy}, {Meibom}, {Kjeldsen},
  {Lissauer}, {Monet}, {Morrison}, {Sasselov}, {Tarter}, {Boss}, {Brownlee},
  {Owen}, {Buzasi}, {Charbonneau}, {Doyle}, {Fortney}, {Ford}, {Holman},
  {Seager}, {Steffen}, {Welsh}, {Rowe}, {Anderson}, {Buchhave}, {Ciardi},
  {Walkowicz}, {Sherry}, {Horch}, {Isaacson}, {Everett}, {Fischer}, {Torres},
  {Johnson}, {Endl}, {MacQueen}, {Bryson}, {Dotson}, {Haas}, {Kolodziejczak},
  {Van Cleve}, {Chandrasekaran}, {Twicken}, {Quintana}, {Clarke}, {Allen},
  {Li}, {Wu}, {Tenenbaum}, {Verner}, {Bruhweiler}, {Barnes}, \&
  {Prsa}}]{Borucki}
{Borucki}, W.~J., {Koch}, D., {Basri}, G., {et~al.} 2010, Science, 327, 977,
  \dodoi{10.1126/science.1185402}

\bibitem[{{Bryson} {et~al.}(2010){Bryson}, {Tenenbaum}, {Jenkins},
  {Chandrasekaran}, {Klaus}, {Caldwell}, {Gilliland}, {Haas}, {Dotson}, {Koch},
  \& {Borucki}}]{Bryson10}
{Bryson}, S.~T., {Tenenbaum}, P., {Jenkins}, J.~M., {et~al.} 2010, \apjl, 713,
  L97, \dodoi{10.1088/2041-8205/713/2/L97}

\bibitem[{{Carry} {et~al.}(2019){Carry}, {Vachier}, {Berthier}, {Marsset},
  {Vernazza}, {Grice}, {Merline}, {Lagadec}, {Fienga}, {Conrad},
  {Podlewska-Gaca}, {Santana-Ros}, {Viikinkoski}, {Hanu{\v{s}}}, {Dumas},
  {Drummond}, {Tamblyn}, {Chapman}, {Behrend}, {Bernasconi}, {Bartczak},
  {Benkhaldoun}, {Birlan}, {Castillo-Rogez}, {Cipriani}, {Colas}, {Drouard},
  {{\v{D}}urech}, {Enke}, {Fauvaud}, {Ferrais}, {Fetick}, {Fusco}, {Gillon},
  {Jehin}, {Jorda}, {Kaasalainen}, {Keppler}, {Kryszczynska}, {Lamy},
  {Marchis}, {Marciniak}, {Michalowski}, {Michel}, {Pajuelo}, {Tanga}, {Vigan},
  {Warner}, {Witasse}, {Yang}, \& {Zurlo}}]{Carry}
{Carry}, B., {Vachier}, F., {Berthier}, J., {et~al.} 2019, \aap, 623, A132,
  \dodoi{10.1051/0004-6361/201833898}

\bibitem[{{Durech} {et~al.}(2011){Durech}, {Kaasalainen}, {Herald}, {Dunham},
  {Timerson}, {Hanu{\v{s}}}, {Frappa}, {Talbot}, {Hayamizu}, {Warner},
  {Pilcher}, \& {Gal{\'a}d}}]{Durech2011}
{Durech}, J., {Kaasalainen}, M., {Herald}, D., {et~al.} 2011, \icarus, 214,
  652, \dodoi{10.1016/j.icarus.2011.03.016}

\bibitem[{Durech \& Kaasalainen(2010)}]{Durech2010}
Durech, J.;~Sidorin, V., \& Kaasalainen, M. 2010, A\&A, 513, A46,
  \dodoi{10.1051/0004-6361/200912693}

\bibitem[{{Gould} \& {Horne}(2013)}]{Gould13}
{Gould}, A., \& {Horne}, K. 2013, \apjl, 779, L28,
  \dodoi{10.1088/2041-8205/779/2/L28}

\bibitem[{{Hanu{\v{s}}} {et~al.}(2018){Hanu{\v{s}}}, {Delbo'}, {{\v{D}}urech},
  \& {Al{\'\i}-Lagoa}}]{Hanus}
{Hanu{\v{s}}}, J., {Delbo'}, M., {{\v{D}}urech}, J., \& {Al{\'\i}-Lagoa}, V.
  2018, \icarus, 309, 297, \dodoi{10.1016/j.icarus.2018.03.016}

\bibitem[{{Henderson} {et~al.}(2016){Henderson}, {Poleski}, {Penny}, {Street},
  {Bennett}, {Hogg}, {Gaudi}, {K2 Campaign 9 Microlensing Science Team}, {Zhu},
  {Barclay}, {Barentsen}, {Howell}, {Mullally}, {Udalski}, {Szyma{\'n}ski},
  {Skowron}, {Mr{\'o}z}, {Koz{\l}owski}, {Wyrzykowski}, {Pietrukowicz},
  {Soszy{\'n}ski}, {Ulaczyk}, {Pawlak}, {OGLE Project}, {Sumi}, {Abe},
  {Asakura}, {Barry}, {Bhattacharya}, {Bond}, {Donachie}, {Freeman}, {Fukui},
  {Hirao}, {Itow}, {Koshimoto}, {Li}, {Ling}, {Masuda}, {Matsubara}, {Muraki},
  {Nagakane}, {Ohnishi}, {Oyokawa}, {Rattenbury}, {Saito}, {Sharan},
  {Sullivan}, {Tristram}, {Yonehara}, {MOA Collaboration}, {Bachelet},
  {Bramich}, {Cassan}, {Dominik}, {Figuera Jaimes}, {Horne}, {Hundertmark},
  {Mao}, {Ranc}, {Schmidt}, {Snodgrass}, {Steele}, {Tsapras}, {Wambsganss},
  {RoboNet Project}, {Burgdorf}, {J{\o}rgensen}, {Calchi Novati}, {Ciceri},
  {D'Ago}, {Evans}, {Hessman}, {Hinse}, {Husser}, {Mancini}, {Popovas},
  {Rabus}, {Rahvar}, {Scarpetta}, {Skottfelt}, {Southworth}, {Unda-Sanzana},
  {MiNDSTEp Team}, {Bryson}, {Caldwell}, {Haas}, {Larson}, {McCalmont},
  {Packard}, {Peterson}, {Putnam}, {Reedy}, {Ross}, {Van Cleve}, {K2C9
  Engineering Team}, {Akeson}, {Batista}, {Beaulieu}, {Beichman}, {Bryden},
  {Ciardi}, {Cole}, {Coutures}, {Foreman-Mackey}, {Fouqu{\'e}}, {Friedmann},
  {Gelino}, {Kaspi}, {Kerins}, {Korhonen}, {Lang}, {Lee}, {Lineweaver}, {Maoz},
  {Marquette}, {Mogavero}, {Morales}, {Nataf}, {Pogge}, {Santerne},
  {Shvartzvald}, {Suzuki}, {Tamura}, {Tisserand}, \& {Wang}}]{Henderson16}
{Henderson}, C.~B., {Poleski}, R., {Penny}, M., {et~al.} 2016, \pasp, 128,
  124401, \dodoi{10.1088/1538-3873/128/970/124401}

\bibitem[{{Holsapple}(2007)}]{Holsapple}
{Holsapple}, K.~A. 2007, \icarus, 187, 500,
  \dodoi{10.1016/j.icarus.2006.08.012}

\bibitem[{{Howell} {et~al.}(2014){Howell}, {Sobeck}, {Haas}, {Still},
  {Barclay}, {Mullally}, {Troeltzsch}, {Aigrain}, {Bryson}, {Caldwell},
  {Chaplin}, {Cochran}, {Huber}, {Marcy}, {Miglio}, {Najita}, {Smith},
  {Twicken}, \& {Fortney}}]{Howell}
{Howell}, S.~B., {Sobeck}, C., {Haas}, M., {et~al.} 2014, \pasp, 126, 398,
  \dodoi{10.1086/676406}

\bibitem[{{Kaasalainen} \& {Torppa}(2001)}]{Kas}
{Kaasalainen}, M., \& {Torppa}, J. 2001, \icarus, 153, 24,
  \dodoi{10.1006/icar.2001.6673}

\bibitem[{{Kalup} {et~al.}(2021){Kalup}, {Moln{\'a}r}, {Kiss}, {Szab{\'o}},
  {P{\'a}l}, {Szak{\'a}ts}, {S{\'a}rneczky}, {Vink{\'o}}, {Szab{\'o}},
  {Kecskem{\'e}thy}, \& {Kiss}}]{Kalup}
{Kalup}, C.~E., {Moln{\'a}r}, L., {Kiss}, C., {et~al.} 2021, arXiv e-prints,
  arXiv:2102.09447.
\newblock \doarXiv{2102.09447}

\bibitem[{{Kurtz}(1985)}]{Kurtz85}
{Kurtz}, D.~W. 1985, \mnras, 213, 773, \dodoi{10.1093/mnras/213.4.773}

\bibitem[{{Lauretta} {et~al.}(2017){Lauretta}, {Balram-Knutson}, {Beshore},
  {Boynton}, {Drouet d'Aubigny}, {DellaGiustina}, {Enos}, {Golish},
  {Hergenrother}, {Howell}, {Bennett}, {Morton}, {Nolan}, {Rizk}, {Roper},
  {Bartels}, {Bos}, {Dworkin}, {Highsmith}, {Lorenz}, {Lim}, {Mink}, {Moreau},
  {Nuth}, {Reuter}, {Simon}, {Bierhaus}, {Bryan}, {Ballouz}, {Barnouin},
  {Binzel}, {Bottke}, {Hamilton}, {Walsh}, {Chesley}, {Christensen}, {Clark},
  {Connolly}, {Crombie}, {Daly}, {Emery}, {McCoy}, {McMahon}, {Scheeres},
  {Messenger}, {Nakamura-Messenger}, {Righter}, \& {Sandford}}]{OSIRIS}
{Lauretta}, D.~S., {Balram-Knutson}, S.~S., {Beshore}, E., {et~al.} 2017, \ssr,
  212, 925, \dodoi{10.1007/s11214-017-0405-1}

\bibitem[{{Marton} {et~al.}(2020){Marton}, {Kiss}, {Moln{\'a}r}, {P{\'a}l},
  {Farkas-Tak{\'a}cs}, {Szab{\'o}}, {M{\"u}ller}, {Ali-Lagoa}, {Szab{\'o}},
  {Vink{\'o}}, {S{\'a}rneczky}, {Kalup}, {Marciniak}, {Duffard}, \&
  {Kiss}}]{Marton20}
{Marton}, G., {Kiss}, C., {Moln{\'a}r}, L., {et~al.} 2020, \icarus, 345,
  113721, \dodoi{10.1016/j.icarus.2020.113721}

\bibitem[{{Moln{\'a}r} {et~al.}(2018){Moln{\'a}r}, {P{\'a}l}, {S{\'a}rneczky},
  {Szab{\'o}}, {Vink{\'o}}, {Szab{\'o}}, {Kiss}, {Hanyecz}, {Marton}, \&
  {Kiss}}]{Molnar18}
{Moln{\'a}r}, L., {P{\'a}l}, A., {S{\'a}rneczky}, K., {et~al.} 2018, \apjs,
  234, 37, \dodoi{10.3847/1538-4365/aaa1a1}

\bibitem[{{P{\'a}l} {et~al.}(2018){P{\'a}l}, {Moln{\'a}r}, \& {Kiss}}]{Pal18}
{P{\'a}l}, A., {Moln{\'a}r}, L., \& {Kiss}, C. 2018, \pasp, 130, 114503,
  \dodoi{10.1088/1538-3873/aae2aa}

\bibitem[{{P{\'a}l} {et~al.}(2020){P{\'a}l}, {Szak{\'a}ts}, {Kiss}, {B{\'o}di},
  {Bogn{\'a}r}, {Kalup}, {Kiss}, {Marton}, {Moln{\'a}r}, {Plachy},
  {S{\'a}rneczky}, {Szab{\'o}}, \& {Szab{\'o}}}]{Pal_tess}
{P{\'a}l}, A., {Szak{\'a}ts}, R., {Kiss}, C., {et~al.} 2020, \apjs, 247, 26,
  \dodoi{10.3847/1538-4365/ab64f0}

\bibitem[{{Podlewska-Gaca} {et~al.}(2020){Podlewska-Gaca}, {Marciniak},
  {Al{\'\i}-Lagoa}, {Bartczak}, {M{\"u}ller}, {Szak{\'a}ts}, {Duffard},
  {Moln{\'a}r}, {P{\'a}l}, {Butkiewicz-B{\k{a}}k}, {Dudzi{\'n}ski}, {Dziadura},
  {Antonini}, {Asenjo}, {Audejean}, {Benkhaldoun}, {Behrend}, {Bernasconi},
  {Bosch}, {Chapman}, {Dintinjana}, {Farkas}, {Ferrais}, {Geier}, {Grice},
  {Hirsh}, {Jacquinot}, {Jehin}, {Jones}, {Molina}, {Morales}, {Parley},
  {Poncy}, {Roy}, {Santana-Ros}, {Seli}, {Sobkowiak}, {Vereb{\'e}lyi}, \&
  {{\.Z}ukowski}}]{paperI}
{Podlewska-Gaca}, E., {Marciniak}, A., {Al{\'\i}-Lagoa}, V., {et~al.} 2020,
  \aap, 638, A11, \dodoi{10.1051/0004-6361/201936380}

\bibitem[{{Poleski} {et~al.}(2019){Poleski}, {Penny}, {Gaudi}, {Udalski},
  {Ranc}, {Barentsen}, \& {Gould}}]{Poleski_method}
{Poleski}, R., {Penny}, M., {Gaudi}, B.~S., {et~al.} 2019, \aap, 627, A54,
  \dodoi{10.1051/0004-6361/201834544}

\bibitem[{{Pravec} {et~al.}(2000){Pravec}, {Hergenrother}, {Whiteley},
  {{\v{S}}arounov{\'a}}, {Ku{\v{s}}nir{\'a}k}, \& {Wolf}}]{Pravec}
{Pravec}, P., {Hergenrother}, C., {Whiteley}, R., {et~al.} 2000, \icarus, 147,
  477, \dodoi{10.1006/icar.2000.6458}

\bibitem[{{Ricker} {et~al.}(2015){Ricker}, {Winn}, {Vanderspek}, {Latham},
  {Bakos}, {Bean}, {Berta-Thompson}, {Brown}, {Buchhave}, {Butler}, {Butler},
  {Chaplin}, {Charbonneau}, {Christensen-Dalsgaard}, {Clampin}, {Deming},
  {Doty}, {De Lee}, {Dressing}, {Dunham}, {Endl}, {Fressin}, {Ge}, {Henning},
  {Holman}, {Howard}, {Ida}, {Jenkins}, {Jernigan}, {Johnson}, {Kaltenegger},
  {Kawai}, {Kjeldsen}, {Laughlin}, {Levine}, {Lin}, {Lissauer}, {MacQueen},
  {Marcy}, {McCullough}, {Morton}, {Narita}, {Paegert}, {Palle}, {Pepe},
  {Pepper}, {Quirrenbach}, {Rinehart}, {Sasselov}, {Sato}, {Seager},
  {Sozzetti}, {Stassun}, {Sullivan}, {Szentgyorgyi}, {Torres}, {Udry}, \&
  {Villasenor}}]{Ricker}
{Ricker}, G.~R., {Winn}, J.~N., {Vanderspek}, R., {et~al.} 2015, Journal of
  Astronomical Telescopes, Instruments, and Systems, 1, 014003,
  \dodoi{10.1117/1.JATIS.1.1.014003}

\bibitem[{Schulz {et~al.}(2012)Schulz, Sierks, Kueppers, \&
  Accomazzo}]{Rosetta}
Schulz, R., Sierks, H., Kueppers, M., \& Accomazzo, A. 2012, Planetary and
  Space Science, 66, 2, \dodoi{10.1016/j.pss.2011.11.013}

\bibitem[{{Schwarzenberg-Czerny}(1991)}]{Czerny1991}
{Schwarzenberg-Czerny}, A. 1991, \mnras, 253, 198,
  \dodoi{10.1093/mnras/253.2.198}

\bibitem[{{Schwarzenberg-Czerny}(1996)}]{Czerny2}
---. 1996, \apjl, 460, L107, \dodoi{10.1086/309985}

\bibitem[{{Shevchenko} {et~al.}(2016){Shevchenko}, {Belskaya}, {Muinonen},
  {Penttil{\"a}}, {Krugly}, {Velichko}, {Chiorny}, {Slyusarev}, {Gaftonyuk}, \&
  {Tereschenko}}]{Shevchenko}
{Shevchenko}, V.~G., {Belskaya}, I.~N., {Muinonen}, K., {et~al.} 2016, \planss,
  123, 101, \dodoi{10.1016/j.pss.2015.11.007}

\bibitem[{{Szab{\'o}} {et~al.}(2020){Szab{\'o}}, {Kiss}, {Szak{\'a}ts},
  {P{\'a}l}, {Moln{\'a}r}, {S{\'a}rneczky}, {Vink{\'o}}, {Szab{\'o}}, {Marton},
  \& {Kiss}}]{Szabo20}
{Szab{\'o}}, G.~M., {Kiss}, C., {Szak{\'a}ts}, R., {et~al.} 2020, \apjs, 247,
  34, \dodoi{10.3847/1538-4365/ab6b23}

\bibitem[{{Szab{\'o}} {et~al.}(2015){Szab{\'o}}, {S{\'a}rneczky}, {Szab{\'o}},
  {P{\'a}l}, {Kiss}, {Cs{\'a}k}, {Ill{\'e}s}, {R{\'a}cz}, \& {Kiss}}]{Szabo15}
{Szab{\'o}}, R., {S{\'a}rneczky}, K., {Szab{\'o}}, G.~M., {et~al.} 2015, \aj,
  149, 112, \dodoi{10.1088/0004-6256/149/3/112}

\bibitem[{{Szab{\'o}} {et~al.}(2016){Szab{\'o}}, {P{\'a}l}, {S{\'a}rneczky},
  {Szab{\'o}}, {Moln{\'a}r}, {Kiss}, {Hanyecz}, {Plachy}, \& {Kiss}}]{Szabo16}
{Szab{\'o}}, R., {P{\'a}l}, A., {S{\'a}rneczky}, K., {et~al.} 2016, \aap, 596,
  A40, \dodoi{10.1051/0004-6361/201629059}

\bibitem[{{Szak{\'a}ts} {et~al.}(2020){Szak{\'a}ts}, {Kiss},
  {Farkas-Tak{\'a}cs}, {Marton}, {M{\"u}ller}, \& {P{\'a}l}}]{Szakats20}
{Szak{\'a}ts}, R., {Kiss}, C., {Farkas-Tak{\'a}cs}, A., {et~al.} 2020, EPSC2020

\bibitem[{{Vernazza} {et~al.}(2018){Vernazza}, {Bro{\v{z}}}, {Drouard},
  {Hanu{\v{s}}}, {Viikinkoski}, {Marsset}, {Jorda}, {Fetick}, {Carry},
  {Marchis}, {Birlan}, {Fusco}, {Santana-Ros}, {Podlewska-Gaca}, {Jehin},
  {Ferrais}, {Bartczak}, {Dudzi{\'n}ski}, {Berthier}, {Castillo-Rogez},
  {Cipriani}, {Colas}, {Dumas}, {{\v{D}}urech}, {Kaasalainen}, {Kryszczynska},
  {Lamy}, {Le Coroller}, {Marciniak}, {Michalowski}, {Michel}, {Pajuelo},
  {Tanga}, {Vachier}, {Vigan}, {Warner}, {Witasse}, {Yang}, {Asphaug},
  {Richardson}, {{\v{S}}eve{\v{c}}ek}, {Gillon}, \& {Benkhaldoun}}]{Vernazza}
{Vernazza}, P., {Bro{\v{z}}}, M., {Drouard}, A., {et~al.} 2018, \aap, 618,
  A154, \dodoi{10.1051/0004-6361/201833477}

\bibitem[{{Wang} {et~al.}(2016){Wang}, {Hogg}, {Foreman-Mackey}, \&
  {Sch{\"o}lkopf}}]{Wang16}
{Wang}, D., {Hogg}, D.~W., {Foreman-Mackey}, D., \& {Sch{\"o}lkopf}, B. 2016,
  \pasp, 128, 094503, \dodoi{10.1088/1538-3873/128/967/094503}

\bibitem[{{Warner} {et~al.}(2009){Warner}, {Harris}, \& {Pravec}}]{Warner}
{Warner}, B.~D., {Harris}, A.~W., \& {Pravec}, P. 2009, \icarus, 202, 134,
  \dodoi{10.1016/j.icarus.2009.02.003}

\end{thebibliography}
\bibliographystyle{aasjournal}

\end{document}